\documentclass[usenatbib]{mn2e}

\usepackage{graphicx}	
\usepackage{amssymb}
\usepackage[fleqn]{amsmath}
\newcommand{\epseri}{$\epsilon$ Eri}
\newcommand{\ujypbm}{$\mu$Jy\,beam$^{-1}$}

\def\jnl@style{}
\def\aaref@jnl#1{{\jnl@style#1}}

\def\aaref@jnl#1{{\jnl@style#1}}

\def\aj{\aaref@jnl{AJ}}                   
\def\araa{\aaref@jnl{ARA\&A}}             
\def\apj{\aaref@jnl{ApJ}}                 
\def\apjl{\aaref@jnl{ApJ}}                
\def\apjs{\aaref@jnl{ApJS}}               
\def\ao{\aaref@jnl{Appl.~Opt.}}           
\def\apss{\aaref@jnl{Ap\&SS}}             
\def\aap{\aaref@jnl{A\&A}}                
\def\aapr{\aaref@jnl{A\&A~Rev.}}          
\def\aaps{\aaref@jnl{A\&AS}}              
\def\azh{\aaref@jnl{AZh}}                 
\def\baas{\aaref@jnl{BAAS}}               
\def\jrasc{\aaref@jnl{JRASC}}             
\def\memras{\aaref@jnl{MmRAS}}            
\def\mnras{\aaref@jnl{MNRAS}}             
\def\pra{\aaref@jnl{Phys.~Rev.~A}}        
\def\prb{\aaref@jnl{Phys.~Rev.~B}}        
\def\prc{\aaref@jnl{Phys.~Rev.~C}}        
\def\prd{\aaref@jnl{Phys.~Rev.~D}}        
\def\pre{\aaref@jnl{Phys.~Rev.~E}}        
\def\prl{\aaref@jnl{Phys.~Rev.~Lett.}}    
\def\pasp{\aaref@jnl{PASP}}               
\def\pasj{\aaref@jnl{PASJ}}               
\def\qjras{\aaref@jnl{QJRAS}}             
\def\skytel{\aaref@jnl{S\&T}}             
\def\solphys{\aaref@jnl{Sol.~Phys.}}      
\def\sovast{\aaref@jnl{Soviet~Ast.}}      
\def\ssr{\aaref@jnl{Space~Sci.~Rev.}}     
\def\zap{\aaref@jnl{ZAp}}                 
\def\nat{\aaref@jnl{Nature}}              
\def\iaucirc{\aaref@jnl{IAU~Circ.}}       
\def\aplett{\aaref@jnl{Astrophys.~Lett.}} 
\def\apspr{\aaref@jnl{Astrophys.~Space~Phys.~Res.}}
\def\bain{\aaref@jnl{Bull.~Astron.~Inst.~Netherlands}} 
\def\fcp{\aaref@jnl{Fund.~Cosmic~Phys.}}  
\def\gca{\aaref@jnl{Geochim.~Cosmochim.~Acta}}   
\def\grl{\aaref@jnl{Geophys.~Res.~Lett.}} 
\def\jcp{\aaref@jnl{J.~Chem.~Phys.}}      
\def\jgr{\aaref@jnl{J.~Geophys.~Res.}}    
\def\jqsrt{\aaref@jnl{J.~Quant.~Spec.~Radiat.~Transf.}}
\def\memsai{\aaref@jnl{Mem.~Soc.~Astron.~Italiana}}
\def\nphysa{\aaref@jnl{Nucl.~Phys.~A}}   
\def\physrep{\aaref@jnl{Phys.~Rep.}}   
\def\physscr{\aaref@jnl{Phys.~Scr}}   
\def\planss{\aaref@jnl{Planet.~Space~Sci.}}   
\def\procspie{\aaref@jnl{Proc.~SPIE}}   

\begin{document}

\title[$\epsilon$ Eri's Debris Ring as Seen by ALMA]{The Northern Arc of $\epsilon$ Eridani's Debris Ring as Seen by ALMA}
\author[M. Booth et al.]{Mark Booth$^{1,2}$\thanks{E-mail: markbooth@cantab.net}, William R. F. Dent$^{3}$, Andr\'es Jord\'an$^{2,4}$, Jean-Fran\c{c}ois Lestrade$^{5}$, \newauthor Antonio S. Hales$^{3,6}$, Mark C. Wyatt$^{7}$, Simon Casassus$^{8,9}$, Steve Ertel$^{10}$, \newauthor Jane S. Greaves$^{11}$, Grant M. Kennedy$^{7}$, Luca Matr\`a$^7$, Jean-Charles Augereau$^{12}$ \newauthor  and Eric Villard$^{3}$ \\
$^{1}$ Astrophysikalisches Institut und Universit\"atssternwarte, Friedrich-Schiller-Universit\"at Jena, Schillerg\"a\ss{}chen 2-3, 07745 Jena, \\Germany \\
$^{2}$ Instituto de Astrof\'isica, Pontificia Universidad Cat\'olica de Chile, Vicu\~na Mackenna 4860, Santiago, Chile \\
$^{3}$ Joint ALMA Observatory, Alonso de C\'ordova 3107, Vitacura 763-0355, Santiago, Chile \\
$^{4}$ Millennium Institute of Astrophysics, Vicu\~na Mackenna 4860, Santiago, Chile\\
$^{5}$ Observatoire de Paris, PSL Research University, CNRS, Sorbonne Universit\'es, UPMC, 61 Av. de l'Observatoire, F-75014 Paris, \\France \\
$^{6}$ National Radio Astronomy Observatory, 520 Edgemont Road, Charlottesville, Virginia, 22903-2475, USA \\
$^{7}$ Institute of Astronomy, University of Cambridge, Madingley Road, Cambridge CB3 0HA, UK \\
$^{8}$ Departamento de Astronomia, Universidad de Chile, Casilla 36-D, Santiago, Chile \\
$^{9}$ Millennium Nucleus ``Protoplanetary Disks'', Santiago, Chile \\
$^{10}$ Steward Observatory, Department of Astronomy, University of Arizona, 993 N. Cherry Ave, Tucson, AZ, 85721, USA \\
$^{11}$ School of Physics and Astronomy, Cardiff University, Queen's Buildings, The Parade, Cardiff CF24 3AA, UK \\
$^{12}$ Univ. Grenoble Alpes, CNRS, IPAG, F-38000 Grenoble, France 
}

\date{Accepted 2017 May 2. Received 2017 May 2; in original form 2016 December 10}
\pubyear{2017}

\maketitle

\begin{abstract}
We present the first ALMA observations of the closest known extrasolar debris disc. This disc orbits the star \epseri, a K-type star just 3.2~pc away. Due to the proximity of the star, the entire disc cannot fit within the ALMA field of view. Therefore, the observations have been centred 18\arcsec{} North of the star, providing us with a clear detection of the northern arc of the ring, at a wavelength of 1.3~mm. The observed disc emission is found to be narrow with a width of just 11-13~AU. The fractional disc width we find is comparable to that of the Solar System's Kuiper Belt and makes this one of the narrowest debris discs known. If the inner and outer edges are due to resonances with a planet then this planet likely has a semi-major axis of 48~AU. We find tentative evidence for clumps in the ring, although there is a strong chance that at least one is a background galaxy. We confirm, at much higher significance, the previous detection of an unresolved emission at the star that is above the level of the photosphere and attribute this excess to stellar chromospheric emission.
\end{abstract}


\section{Introduction}
At a distance of 3.2 pc, the K2V star $\epsilon$ Eridani is the closest extrasolar system known to host a 
debris disc -- a circumstellar disc of dust and planetesimals. This proximity has resulted in it being one of the most intensely studied debris discs since the spatial resolution achievable allows for a detailed analysis of the radial and azimuthal structure of the disc. This is particularly interesting as such analysis can be used to predict the presence of planets in a system \citep[see e.g.][]{moro13}. 

\epseri{} was one of the first four debris discs to be discovered \citep{gillett86}. Early submillimetre and millimetre observations suggested that it was slightly extended compared to the beam \citep{chini91}. With the introduction of the Submillimetre Common-User Bolometer Array (SCUBA) to the James Clerk Maxwell Telescope (JCMT), \citet{greaves98} were able to image the disc finding it to be a ring at $\sim60$~AU with an orientation close to face-on and resolving the inner edge of a debris disc for the first time. They also noticed `clumpy' structures in the disc, that they proposed could be due to resonant trapping of dust by planets. This inspired a number of authors to numerically determine properties of the planetary system. \citet{ozernoy00} modelled the interaction between a planet at 60~AU and the disc and predicted that if the clumps are created by resonances with a planet they should orbit the star at a rate of 0.6-0.8\degr{}yr$^{-1}$. \citet{quillen02} used the features of the disc to predict that the planet would be on an eccentric orbit with a lower semi-major axis of $\sim$40~AU and this would result in the resonant features orbiting at a faster rate and \citet{deller05} note that at least one extra planet is needed to explain the lack of sub-mm emission closer to the star. Follow-up observations with SCUBA \citep{greaves05} suggest that the clumps appear to be orbiting at a rate of 2.75\degr{}yr$^{-1}$ anti-clockwise \citep{poulton06}, but leave open the possibility that many of the clumps are actually background galaxies. Observations with the Institut de Radioastronomie Millim\'etrique (IRAM) 30m telescope \citep{lestrade15}, the Submillimeter Array (SMA) \citep{macgregor15} and SCUBA-2 on the JCMT \citep{holland17} show some tentative signs of asymmetries, but observations with the Caltech Submillimeter Observatory (CSO) \citep{backman09}, the Herschel Space Observatory \citep{greaves14a} and the Large Millimeter Telescope (LMT) \citep{chavez16} seem to show smoother structure. \citet{greaves14a} and \citet{chavez16} also find that there are a large number of background galaxies to the East of \epseri, some of which would have been behind the disc at the time of the earlier observations due to \epseri's high proper motion of almost 1\arcsec{}yr$^{-1}$ in a westward direction.

In addition to the main belt seen at long wavelengths, the 24-160 $\mu$m spectral energy distribution (SED) from Spitzer suggests a warmer ring
at 20~AU and possibly another at 3~AU \citep{backman09}. \citet{reidemeister11} propose an alternative model with parent planetesimals just located at the main belt, which then produce dust that fills the inner part of the system via drag forces due to the stellar radiation and stellar wind. Observations with Herschel marginally resolve a ring at $\sim$14~AU rather than 20~AU with the same orientation as the outer ring \citep{greaves14a}. Recent observations from the LMT at 1.1mm \citep{chavez16} go deep enough to clearly show dust emission in between the star and the outer ring. The detection of dust interior to the main ring in the mm calls into question the \citet{reidemeister11} model as they show that the drag forces only have a strong effect on grains with sizes $\lesssim 10\,\mu$m that are expected to be very faint at long wavelengths. \citet{su17} show that the \citep{reidemeister11} model is inconsistent with observations from the Stratospheric Observatory for Infrared Astronomy (SOFIA), whilst the models of \citet{backman09} and \citet{greaves14a} are consistent with them.

In this paper we present the first Atacama Large Millimeter/submillimeter Array (ALMA) observations of \epseri{}. This provides us with the highest resolution image of the disc. We use a forward modelling technique to determine the geometric properties of the disc. We compare these results with prior observations and discuss the implications for planets in the system and the properties of the star.

\begin{figure*}
	\centering
	\includegraphics[width=0.33\textwidth]{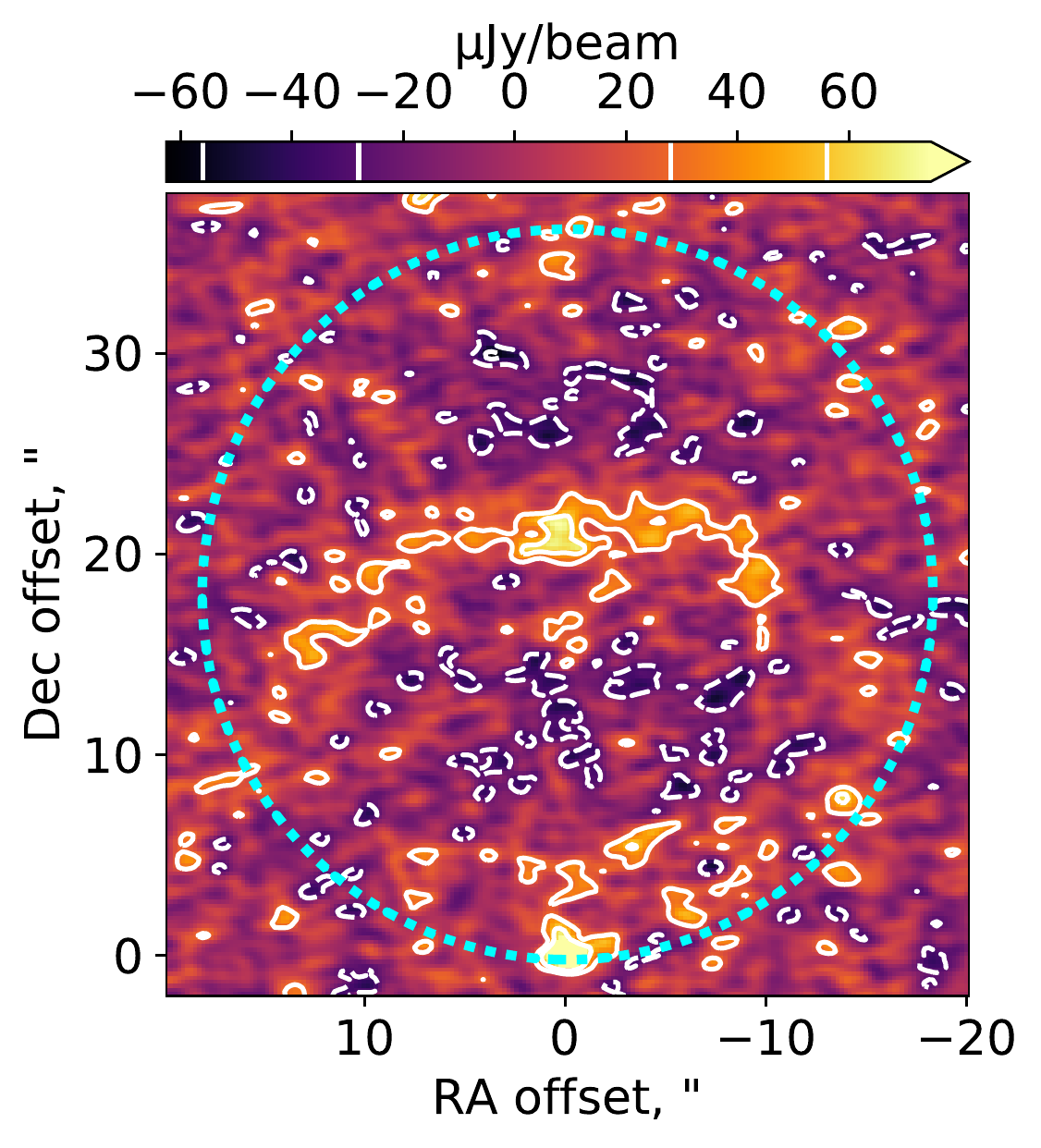}
	\includegraphics[width=0.34\textwidth]{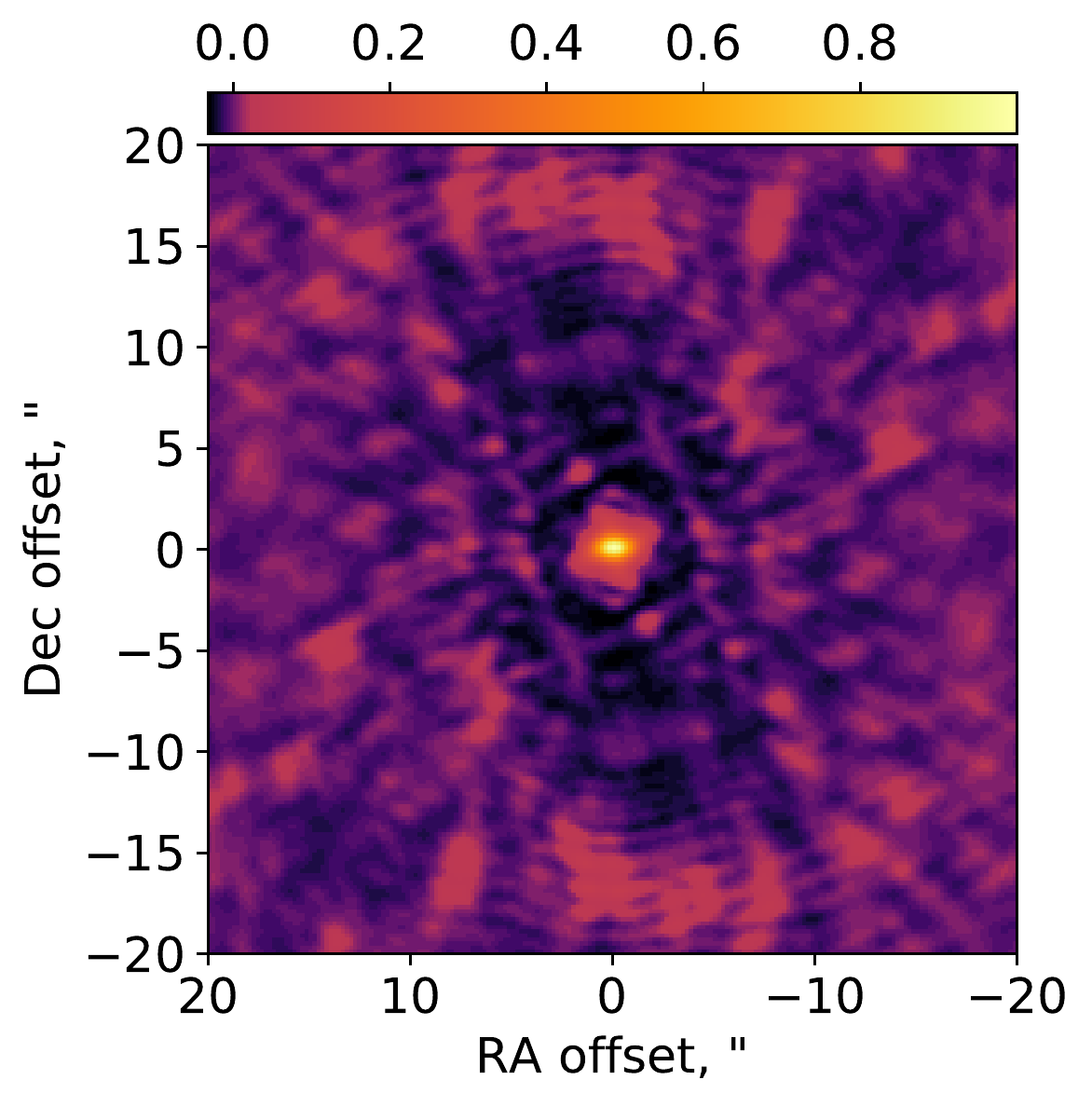}
	\includegraphics[width=0.32\textwidth]{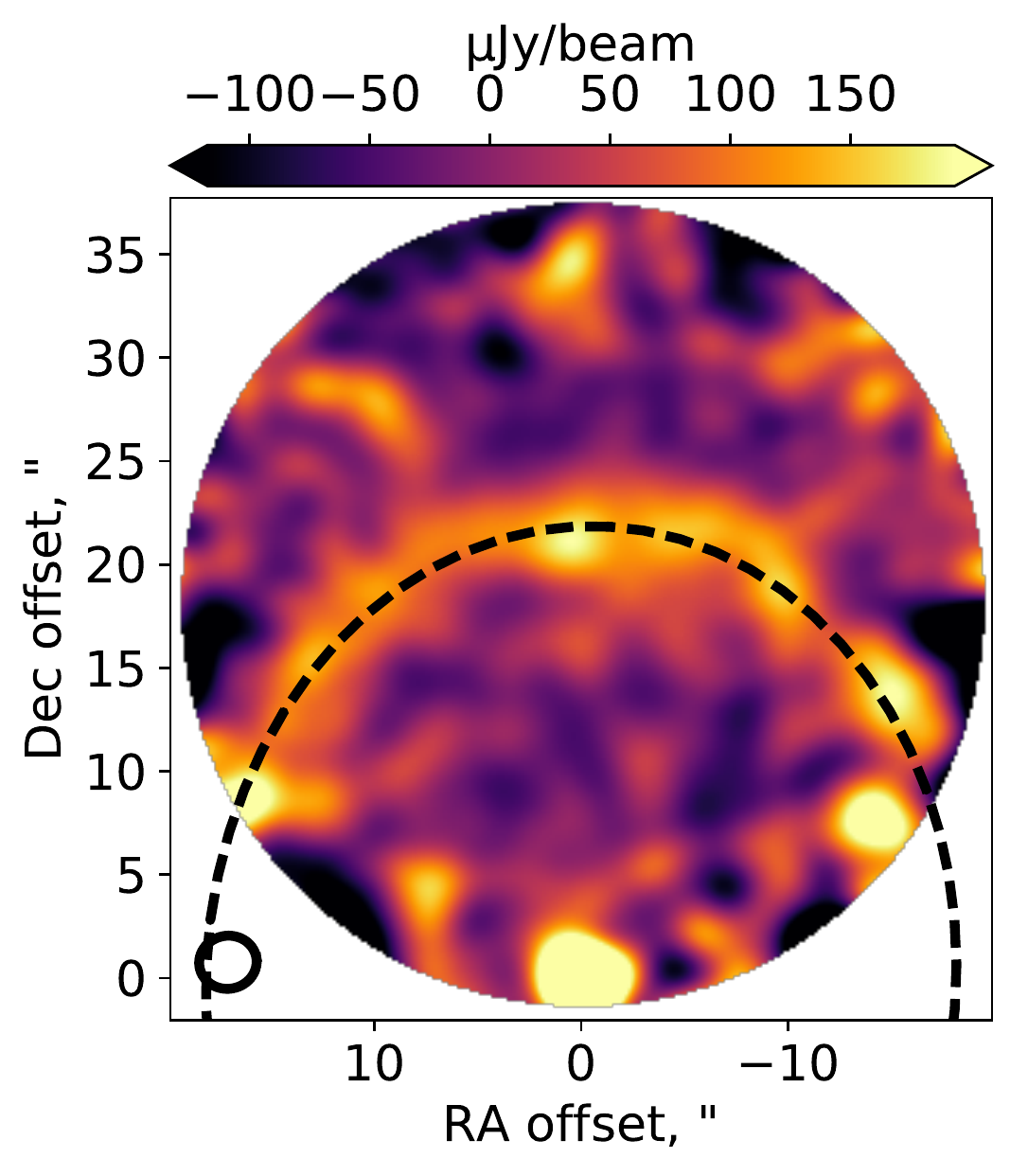}
	\caption{\emph{Left:} Dirty image i.e. the Fourier transform of the visibilities. The star is detected at high significance at the bottom of the image with the ring detected as a faint arc in the centre of the image. Contours are in increments of $\pm2\sigma$. The dotted (cyan) circle illustrates the 25\% primary beam power level. \emph{Middle:} Dirty beam. The low level positive and negative structure is due to the incomplete sampling of the Fourier plane. \emph{Right:} Natural weighted CLEAN image with a taper of 2.5\arcsec{} that has been corrected for the primary beam. The image has been cut-off at the 20\% primary beam power level. The dashed ellipse shows a distance of 70~AU for an inclination of 34\degr{} and position angle of -4\degr.}
	\label{fclean}
\end{figure*}

\section{Observations}
\label{sobs}
\epseri{} was observed by ALMA in cycle 2 as part of the project 2013.1.00645.S (PI: A. Jord\'an). The data were taken in band 6 (1.34~mm) on the 17th and 18th January 2015 (see table \ref{tobs}). The total time on target was 4.4 hours. The proximity of this star to the Earth results in the disc having an angular diameter far larger than the primary beam of ALMA and so we decided to focus our pointing at a position of $\alpha=03:32:54.8535$; $\delta=-09.27.11.438$ (J2000), 18\arcsec{} to the North of the star, coincident with the expected position of the northern ansa of the disc. 

\begin{table}%
\begin{tabular}{cccc}
ID & Observation Start & Antennas & PWV (mm) \\ 
\hline
1 & 2015-01-17 00:37:39 & 34 & 4.0 \\ 
2 & 2015-01-17 20:09:10 & 35 & 4.2 \\ 
3 & 2015-01-17 21:40:16 & 35 & 4.2 \\ 
4 & 2015-01-18 00:12:15 & 35 & 4.2 \\ 
5 & 2015-01-18 02:32:14 & 34 & 4.2 \\ 
6 & 2015-01-18 03:50:01 & 34 & 4.2 \\ 
\end{tabular}
\caption{Details of the observations. PWV stands for precipitable water vapour.}
\label{tobs}
\end{table}

The data were reduced using the standard observatory calibration in \texttt{CASA} version 4.3.1 \citep{mcmullin07}, which includes water vapour radiometry correction, system temperature, complex gain calibration and flagging. The correlator was configured such that one spectral window, centred at the CO(2-1) line at 230.538~GHz, had 3840 channels of width 0.5 MHz (0.6~kms$^{-1}$) and the other three spectral windows, centred at 232, 219 and 217~GHz, had 128 channels of width 16~MHz (21~km\,s$^{-1}$). For the continuum imaging we can, therefore, use the full bandwidth of all four channels adding up to 7.875~GHz. The observations were taken whilst the array was in a compact configuration with baselines between 15 and 350~m.

The dirty image (the inverse Fourier transform of the observed visibilities), created using natural weighting and multi-frequency synthesis, is shown on the left in figure \ref{fclean}. The middle plot of figure \ref{fclean} shows the dirty beam, which is the Fourier transform of the sampling function. The extensive (positive and negative) substructure is due to the lack of short baselines and resulting lack of sensitivity to extended emission. The resulting synthesised beam has a size 1.6\arcsec$\times$1.1{\arcsec} with a position angle of 92\degr{} E of N. As the apparent size of the disc is very large and the dirty beam has extensive substructure, we measure the RMS, $\sigma$, in a region 90\arcsec{} offset from the primary beam to avoid any influence from the emission. We find this to be $\sigma=14$~\ujypbm. Emission is clearly seen where the star is expected to be, despite it being far from the phase centre. The peak flux density is 0.82$\pm$0.07~mJy\footnote{Including a 5\% calibration uncertainty \protect\citep[see section C.4.1][]{alma16}.} after correcting for the primary beam. A Gaussian fit around the location of star shows a peak at $\alpha=03:32:54.861\pm0.004$; $\delta=-09.27.29.415\pm0.003$. This is offset by 0.11\arcsec{} from the expected location, which is consistent with the expected astrometric accuracy of 0.10\arcsec{} \citep[section 10.6.3 of][]{alma16}. The Gaussian fit also shows no significant deviations from the beam shape, i.e. there is no sign of resolved emission at the location of the star.

The debris disc is seen as an arc of mostly low signal-to-noise emission at $\sim$20\arcsec{} from the star, with a peak directly North of the star of 5$\sigma$ significance. 
We also show a CLEAN image\footnote{CLEAN \citep{hogbom74} is an algorithm that fits point sources to the image. The CLEAN image is a restored image created by convolving the CLEAN model with the synthesised beam and adding the residuals.} with a 2.5\arcsec{} taper on the right in figure \ref{fclean} to emphasise the detection at the expense of resolution. This also shows that the peak at the northern ansa is not simply due to the higher sensitivity near to the phase centre. No spectral lines were detected down to an RMS sensitivity per beam of 0.98~mJy\,beam$^{-1}$ in a 1.27~km\,s$^{-1}$ wide channel.

\begin{table*}%
\begin{tabular}{ccccccccc}
&  \multicolumn{2}{c}{Run A} &  \multicolumn{2}{c}{Run B} &  \multicolumn{2}{c}{Run C}\\
Parameter & Prior & Fit & Prior & Fit & Prior & Fit \\ 
\hline
$R_{in}$ (AU)& Uniform & 62.6$^{+0.9}_{-1.5}$ & Uniform  & $<55^a$ & -  & - \\ 
& 57$<R_{in}<$80 & & 0$<R_{in}<$57  & &   & \\
$R_{out}$ (AU) & Uniform & 75.9$^{+1.0}_{-0.9}$ & Uniform  & 75.0$^{+0.7}_{-0.7}$&  - & - \\ 
& $R_{in}<R_{out}<$100 & & $R_{in}<R_{out}<$100  & &   & \\
$\gamma$ & Uniform & 2$^{+3}_{-3}$ & Uniform  & 7$^{+1}_{-1}$& - &  - \\ 
& -8$<\gamma<$10 & & 3$<\gamma<$20  & &   & \\
$R_{mid}$ (AU) & - & - & - & - & Uniform &69.4$^{+0.5}_{-0.4}$\\
&  & &  & & 0$<R_{mid}<$90  & \\
$\Delta R$ (AU)& - & - & - & - & Uniform & 11.3$^{+1.4}_{-1.2}$\\
&  & &   & & 0$<\Delta R<R_{mid}$  & \\
$F_{\nu}$ (mJy) & Uniform in ln & 8.3$^{+0.6}_{-0.6}$ & Uniform in ln  &10$^{+1}_{-1}$ & Uniform in ln  & 9.2$^{+0.8}_{-0.8}$ \\ 
& 1$<\ln(F_{\nu})<$4 & & 1$<\ln(F_{\nu})<$4  & & 1$<\ln(F_{\nu})<$4  & \\
$I$ (\degr)& Uniform in cosine & 34$^{+2}_{-2}$ & Uniform in cosine &33$^{+2}_{-2}$ & Uniform in cosine  & 33$^{+2}_{-2}$ \\ 
& 0$<\cos(I)<$1 & & 0$<\cos(I)<$1  & & 0$<\cos(I)<$1  & \\
$F_{cen}$ (mJy)& Uniform in ln  & 0.86$^{+0.07}_{-0.07}$& Uniform in ln  & 0.82$^{+0.07}_{-0.07}$& Uniform in ln  &0.85$^{+0.07}_{-0.07}$ \\
& -1$<\ln(F_{cen})<$6 & & -1$<\ln(F_{cen})<$6  & & -1$<\ln(F_{cen})<$6  & \\
$\Omega$ (\degr)& Uniform & -4$^{+3}_{-3}$ &  Uniform & -3$^{+3}_{-3}$ & Uniform & -4$^{+3}_{-3}$  \\
& -20$<\Omega<$20 & & -20$<\Omega<$20 & & -20$<\Omega<$20  & \\
\hline
$\chi^2_{red}$ & & 1.100 & & 1.097 & & 1.100 \\
\end{tabular}
\caption{Free parameters, their priors and the results of the fitting. Note that the uncertainties on the flux densities given here do not take into account the 5\% calibration uncertainties. \newline
$^a$ The maximum likelihood occurs for 48~AU, but inner edges as far in as the star still fit the data.}
\label{tfit}
\end{table*}

\begin{figure*}
	\centering
	\includegraphics[width=0.33\textwidth]{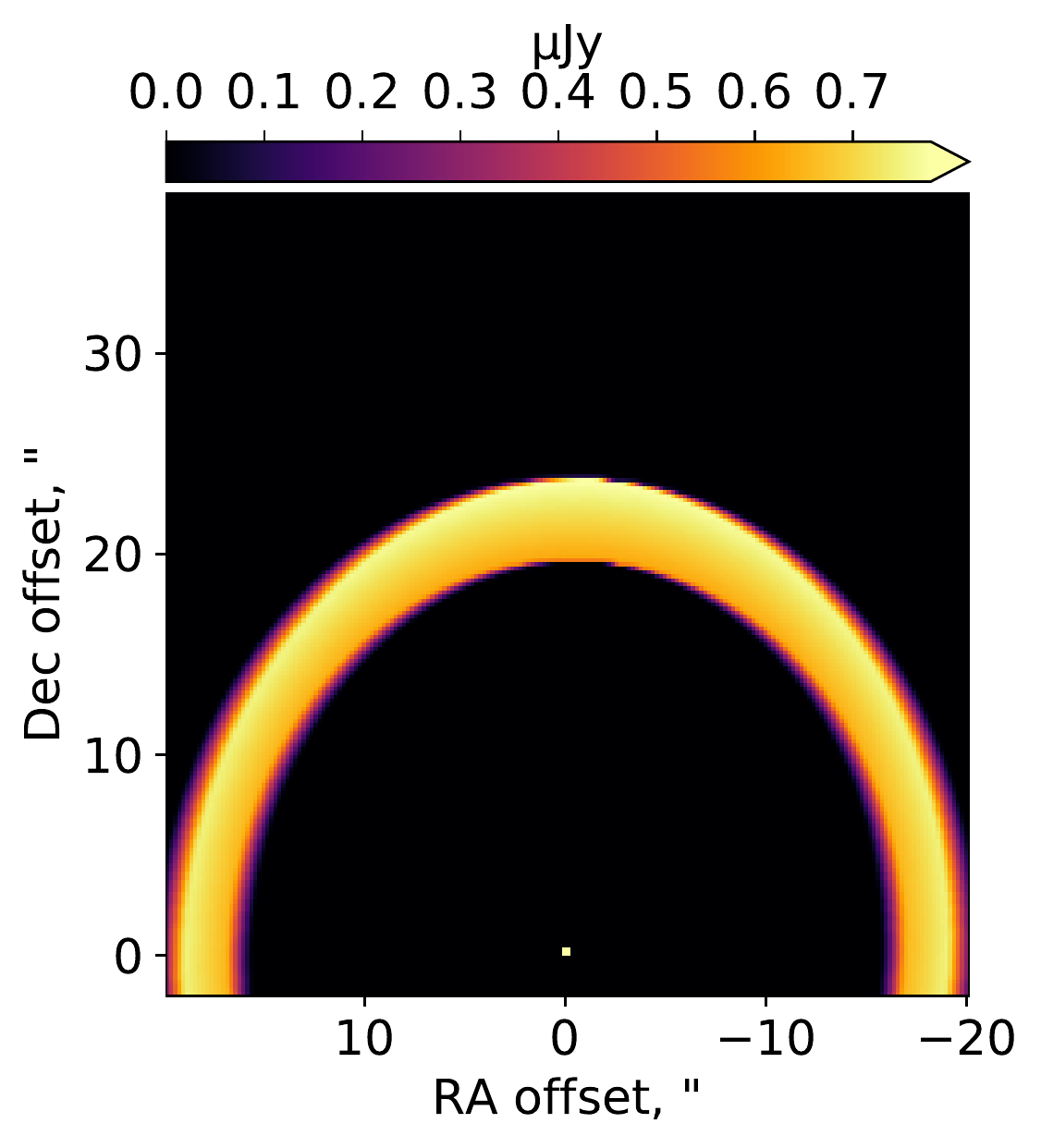}
	\includegraphics[width=0.33\textwidth]{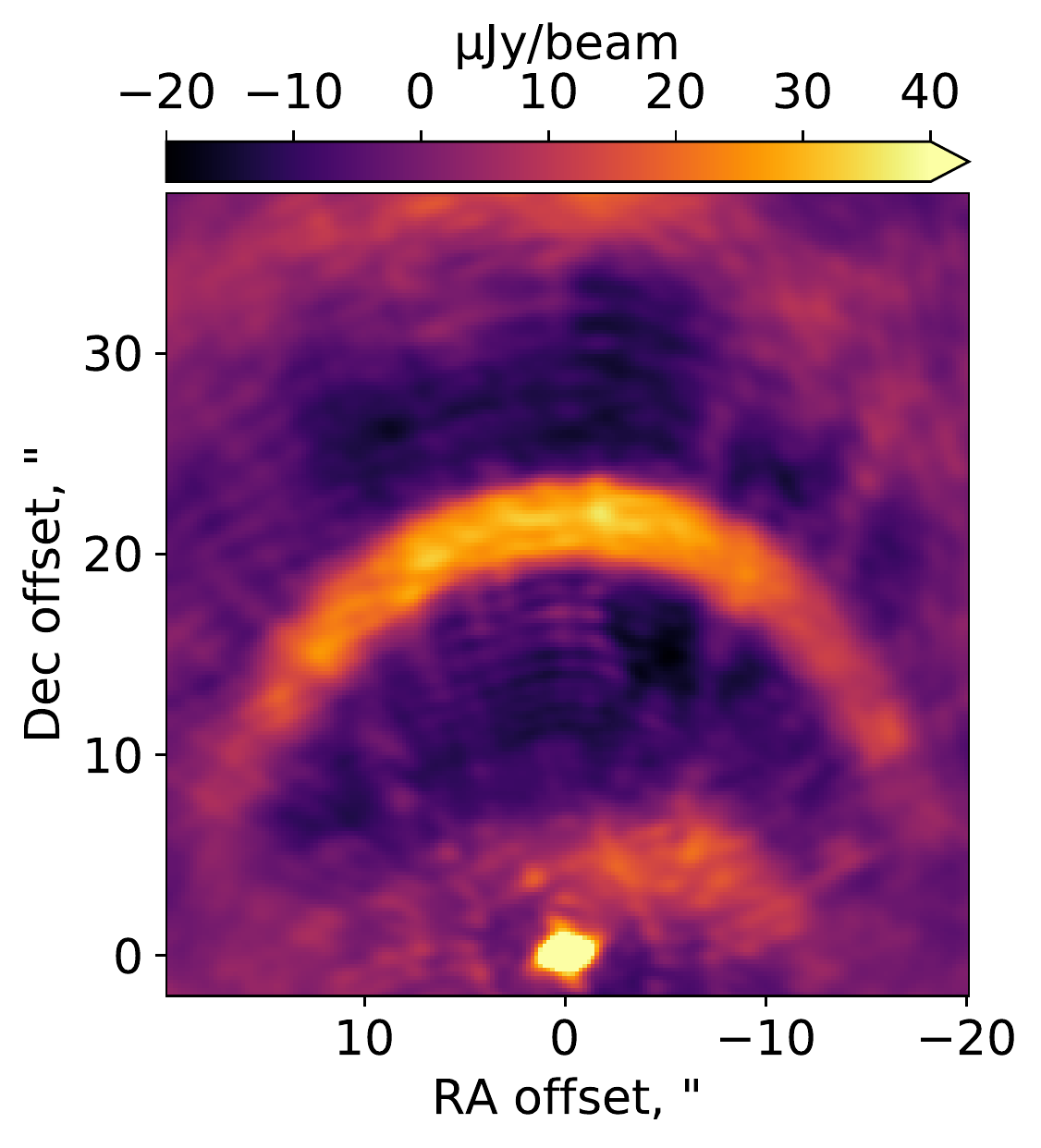}
	\includegraphics[width=0.33\textwidth]{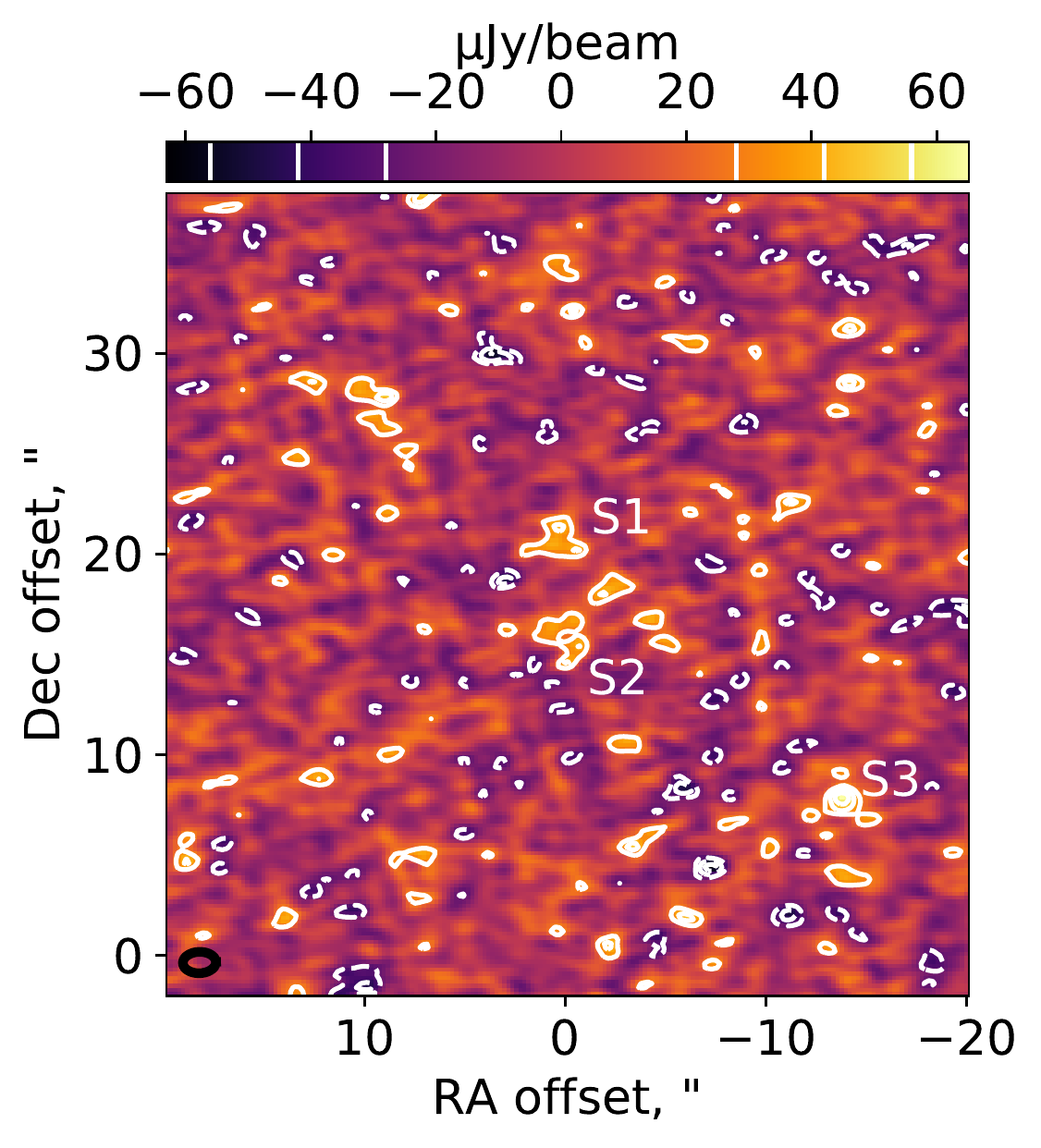} \\
	\includegraphics[width=0.33\textwidth]{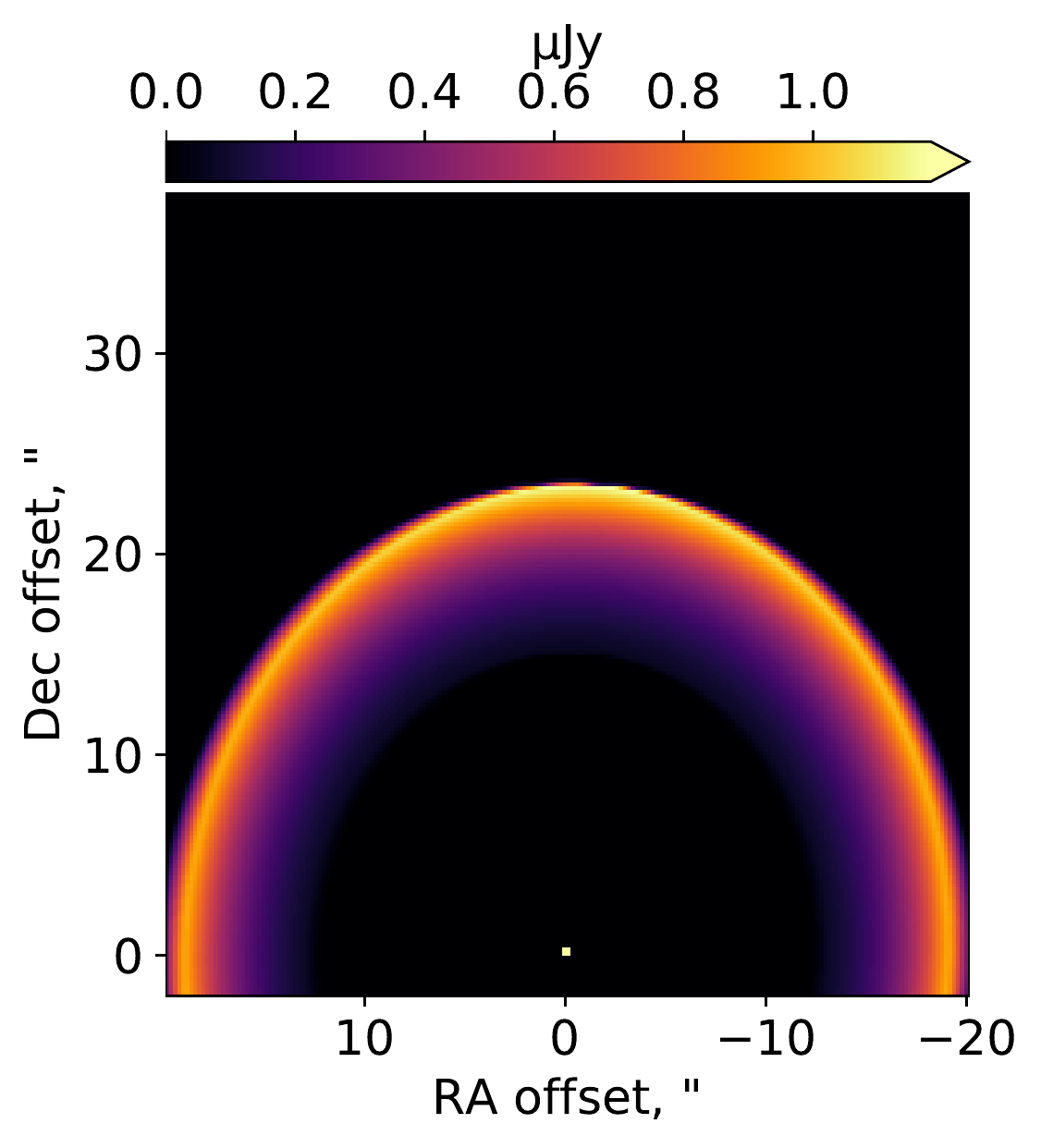}
	\includegraphics[width=0.33\textwidth]{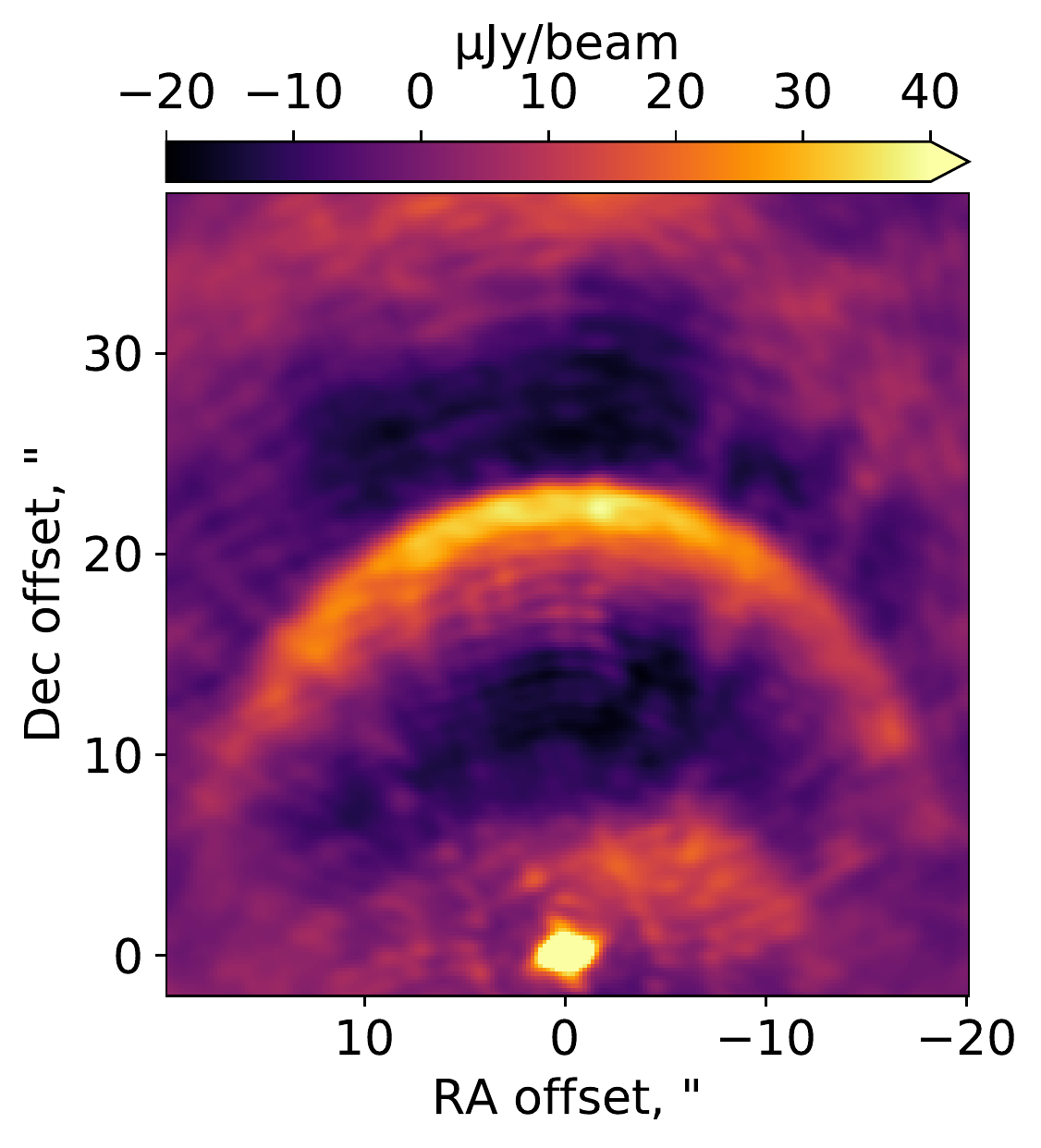}
	\includegraphics[width=0.33\textwidth]{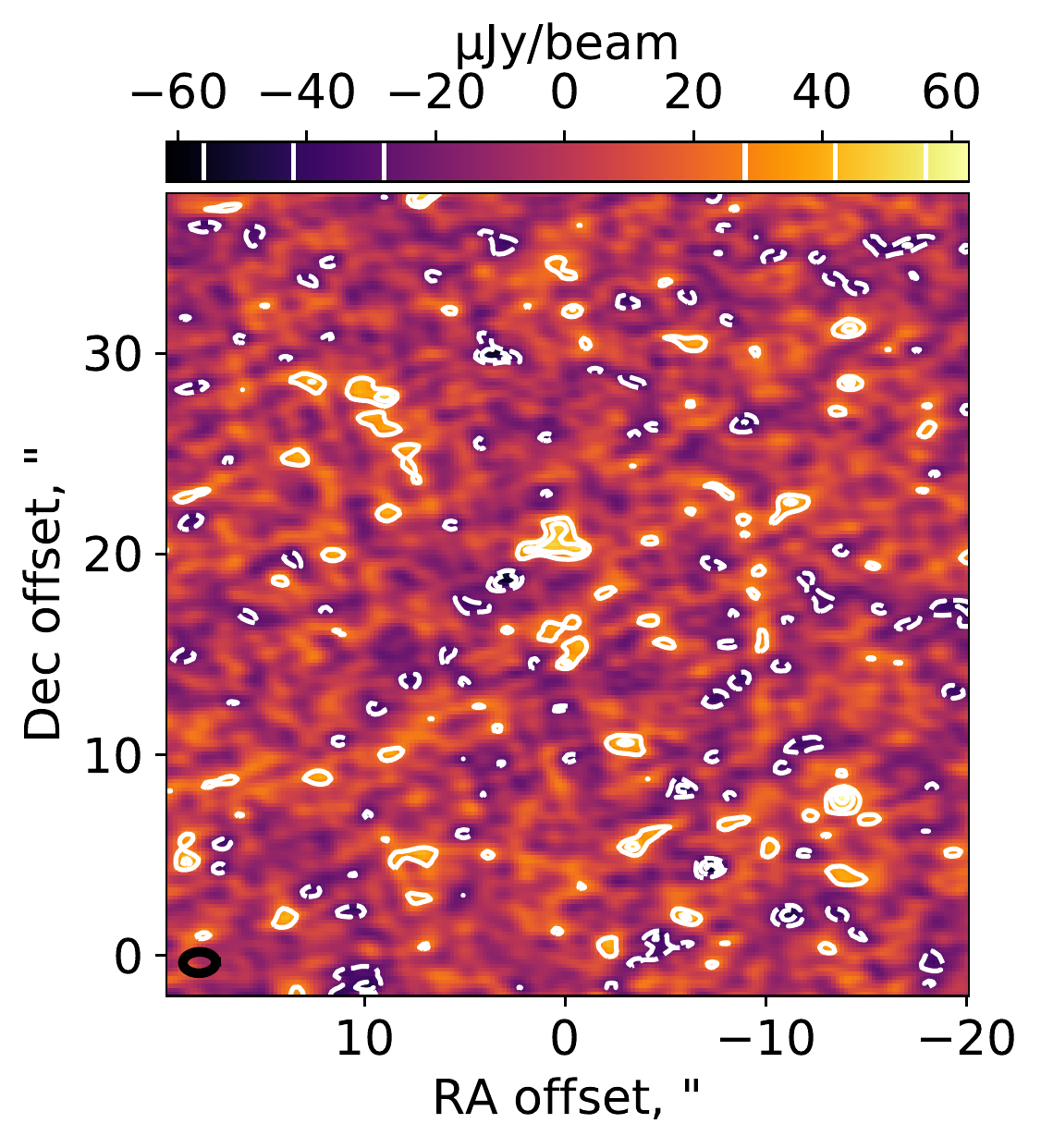} \\
	\includegraphics[width=0.33\textwidth]{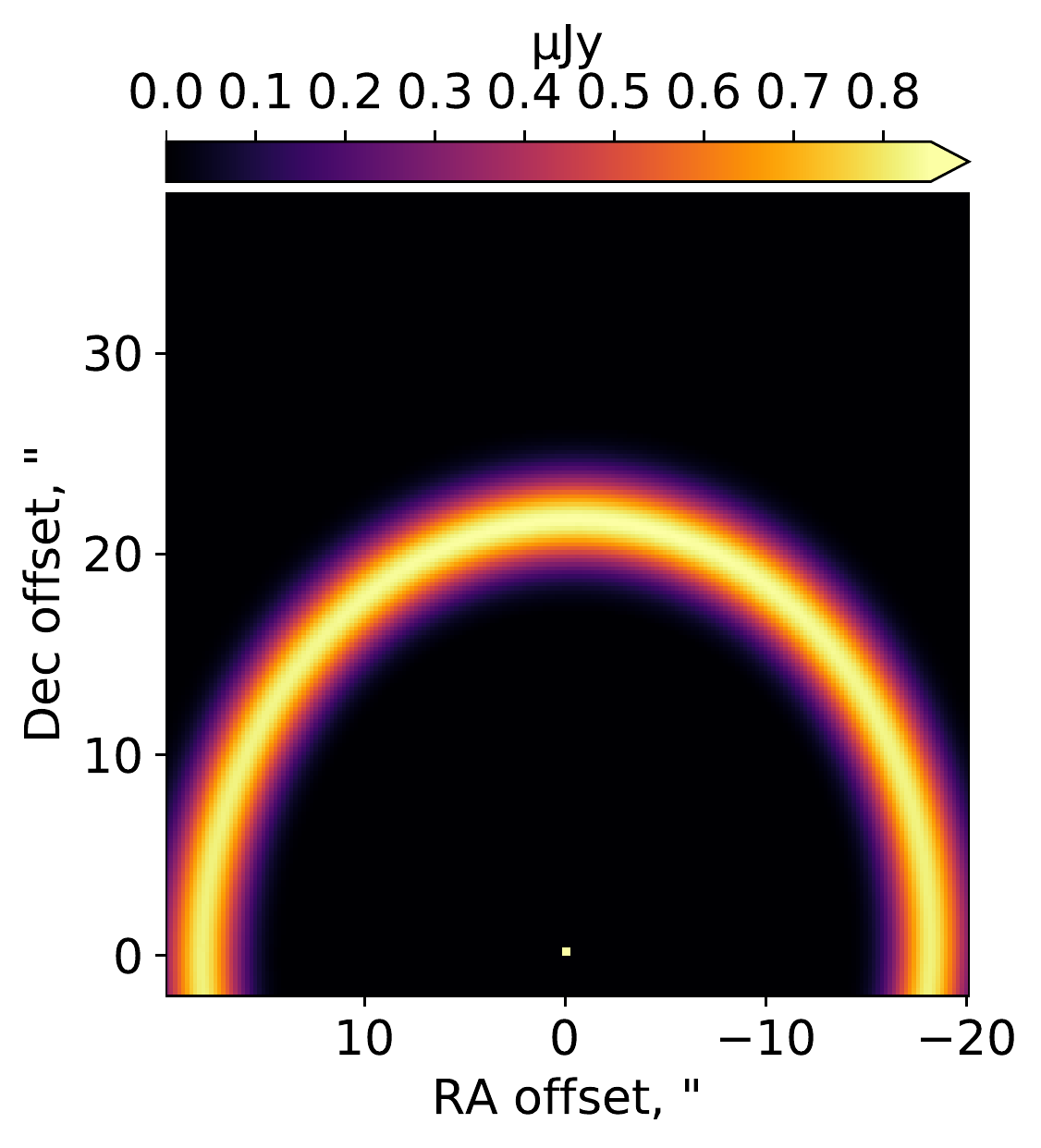}
	\includegraphics[width=0.33\textwidth]{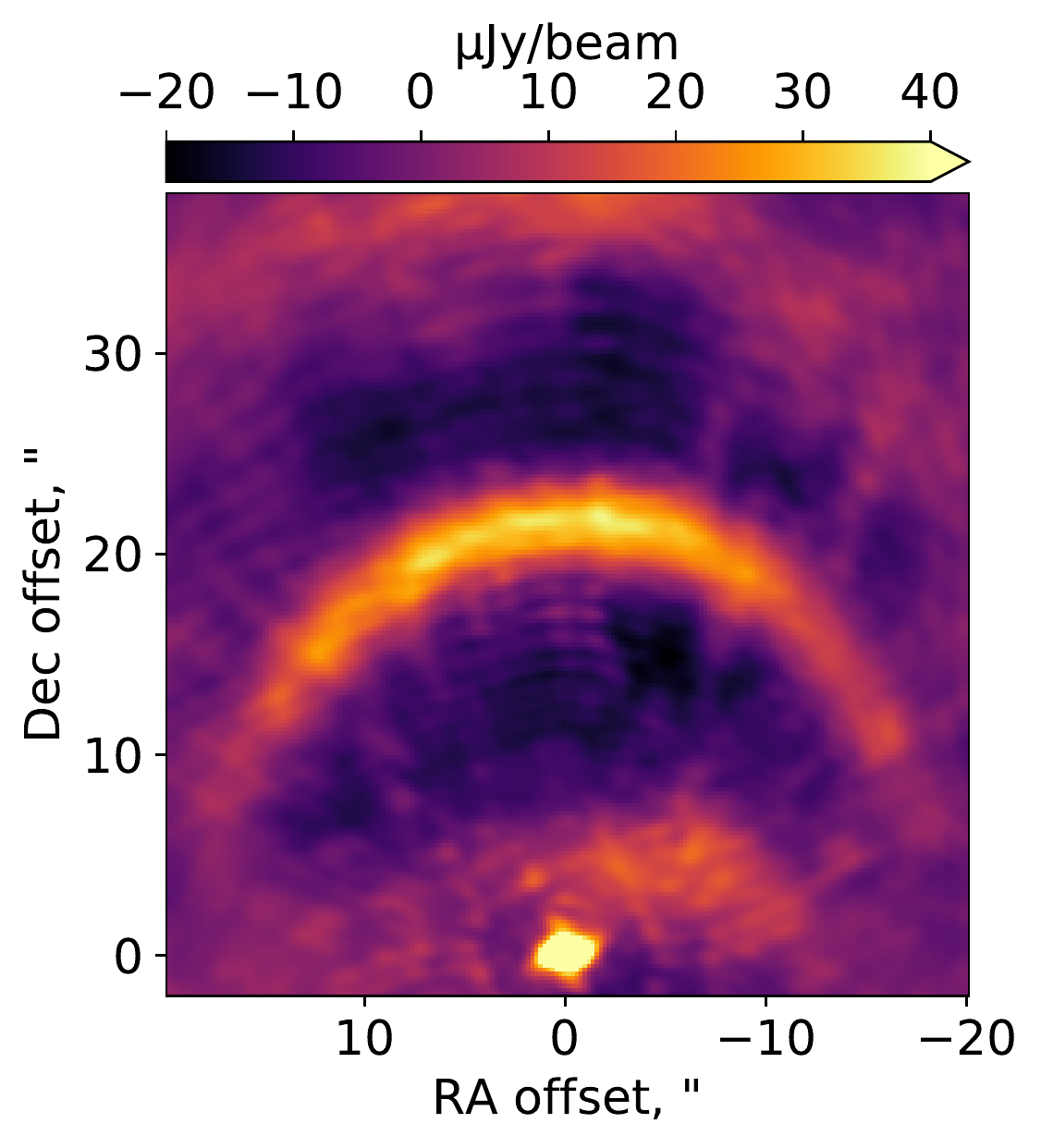}
	\includegraphics[width=0.33\textwidth]{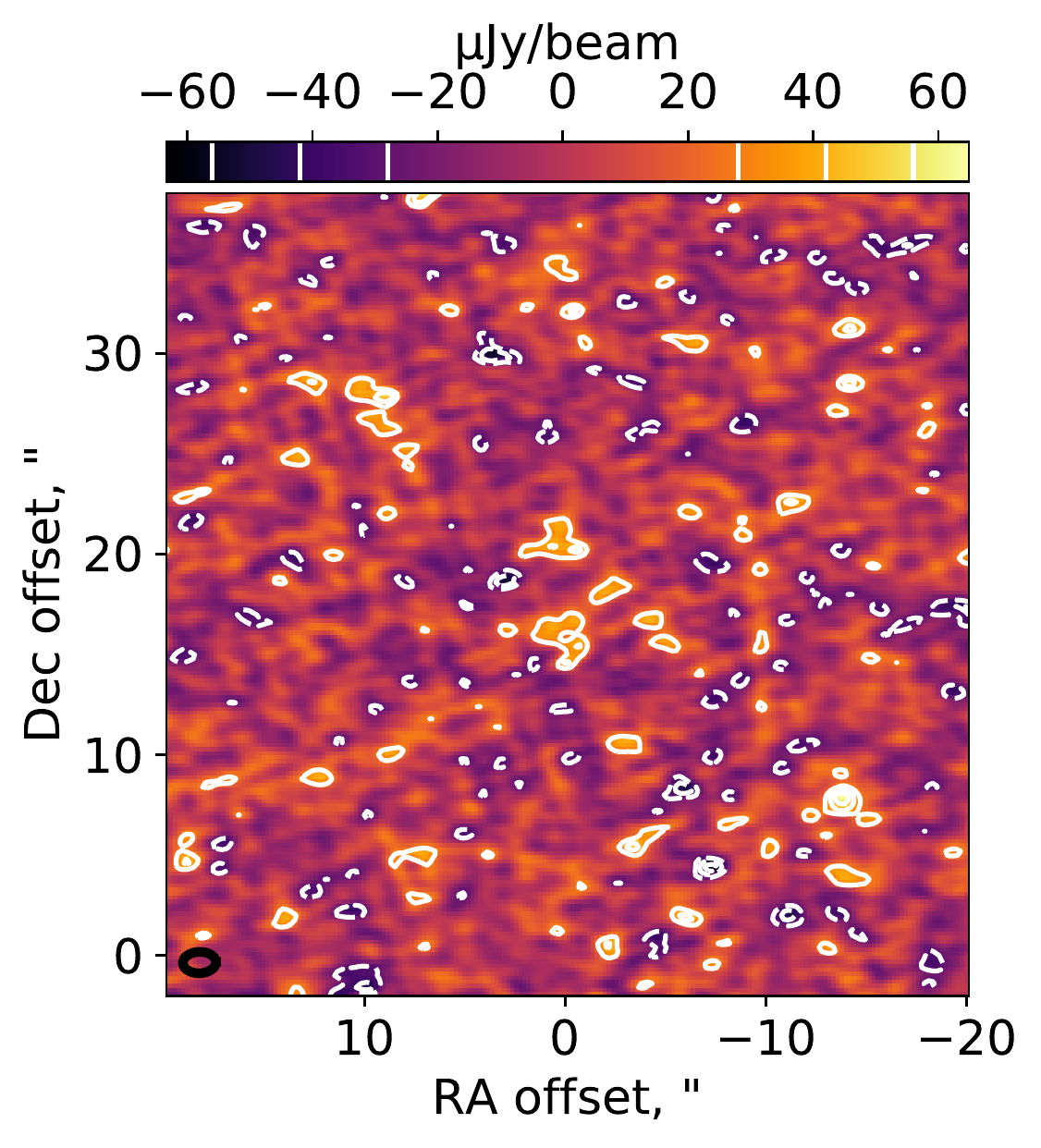}
	\caption{\emph{Left:} Model. \emph{Middle:} Model convolved with the dirty beam and attenuated by the primary beam. \emph{Right:} Residuals after the best-fit model is subtracted from the image. Three of the most significant residuals have been labelled in the top right plot and will be discussed in the text. Contours start at $\pm2\sigma$ and increase in increments of $1\sigma$. \emph{Top row:} Run A. \emph{Middle row:} Run B. \emph{Bottom row:} Run C.}
	\label{fmod}
\end{figure*}

\section{Modelling}
\label{smod}
The modelling procedure used here is based on that of \citet{booth16}. To find the best fit parameters of the disc and their uncertainties, a Markov Chain Monte Carlo (MCMC) routine is run making use of \texttt{emcee} \citep{foreman13}. For each likelihood calculation, a model disc distribution is created using a specified radial distribution (see the following sections) that is inclined from face on by an angle $I$ and given a position angle, $\Omega$, measured anti-clockwise from North. The disc grains are assumed to act like black bodies i.e., they have a temperature that is $\propto R^{-0.5}$. This assumption is made as we are working at a single wavelength, and so the temperature profile is degenerate with the optical depth profile. From the distribution of particles, the sky image is then determined using a line of sight integrator method \citep{wyatt99}. The total flux density of the disc at this wavelength is scaled to $F_\nu$ -- whilst we only observe part of the disc, we model the whole disc and all disc fluxes given in this paper refer to the full ring not just the part visible in the image. A central component, $F_{cen}$, is added to the sky image at the fitted position of the star (see section \ref{sobs}). Since we only observe one side of the disc, it is difficult for us to place any constraints on the centre of the ring and so we assume that it is coincident with the location of the star. Whilst \citet{greaves14a} showed that an offset between the ring centre and the star could explain the brightness asymmetry that they see, no observations have detected any offset yet, with \citet{macgregor15} placing a 3$\sigma$ upper limit on the offset of 2.7\arcsec. The resulting sky image is then multiplied by the primary beam, convolved with the dirty beam and compared with the dirty image to produce a likelihood given by

\begin{equation}
\ln \mathcal{L}=-\chi^2/2
\end{equation}
\begin{equation}
 \chi^2=\sum^{N}_{i=1} \left(\frac{O_i-M_i}{S_{ncr}\sigma}\right)^2
\end{equation}
$O_i$ and $M_i$ represent pixels in the observed image and model image respectively. $N$ is the total number of pixels used in the calculation. We note that the primary beam image we use is a Gaussian model of the beam. The actual primary beam may deviate slightly from this, which adds extra uncertainty to pixels far from the phase centre. We, therefore, only include pixels where the primary beam power is $>$20\% of the peak in the primary beam. $S_{ncr}$ is the noise correlation ratio equivalent to the square root of the number of pixels per beam. This is required because the high resolution of the image compared to the beam size has the side effect of introducing correlated noise.
The MCMC is run with 120 walkers and 2000 timesteps for each model tested.

\subsection{Single component}
\label{ssingle}
Observations at millimetre wavelengths are typically dominated by large grains that are not strongly affected by transport processes and so should be coincident with their `birth-ring'. We, therefore, start by considering a single component disc to fit the main belt. This has sharp edges at $R_{in}$ and $R_{out}$ and an optical depth profile, $\tau$, that varies with radius as $R^\gamma$. Preliminary tests showed good fits in two regions of parameter space, which we, therefore, split into two MCMC runs: run A, which is constrained to be a narrow belt (with $R_{in}>57$~AU), and run B, which is a wide belt ($R_{in}<57$~AU) with a steeply rising profile ($\gamma>3$). The rising slope of the latter run could indicate a low-level effect of drag forces on the grains as in the model of \citet{reidemeister11}. As in \citet{macgregor15}, we also consider a Gaussian function (run C) centred at $R_{mid}$ with a full-width-half-maximum, $\Delta R$, to test whether a smoother edge makes for a better fit. Uniform priors have been assumed in all cases. With the exception of the limits noted above for runs A and B, the limits on the priors have been set to be wide enough so as to not influence the results. The free parameters, their priors and the best fit results are shown in table \ref{tfit}. The reduced $\chi^2$ is given by
\begin{equation}
 \chi_{red}^2=\frac{1}{N}\sum^{N}_{i=1} \left(\frac{O_i-M_i}{\sigma}\right)^2
\end{equation}

The best fit model for each case is shown on the left in figure \ref{fmod}. The middle column of this figure shows the convolved image and the last column shows the residuals for each run. They each show a significant residual roughly coincident with the northern ansa (labelled S1), although for run B it is slightly more significant than runs A and C. There is also another significant residual just to the South of this (S2) and a 4.6$\sigma$ residual in the North-West portion of the disc (S3). These will be discussed further in section \ref{sazi}. Other than this, there is very little difference between the best fits for each of the runs and the $\chi_{red}^2$ values are almost identical meaning that, from the images alone it is not possible to distinguish between these models.

\begin{figure*}
	\centering
	\includegraphics[width=0.33\textwidth]{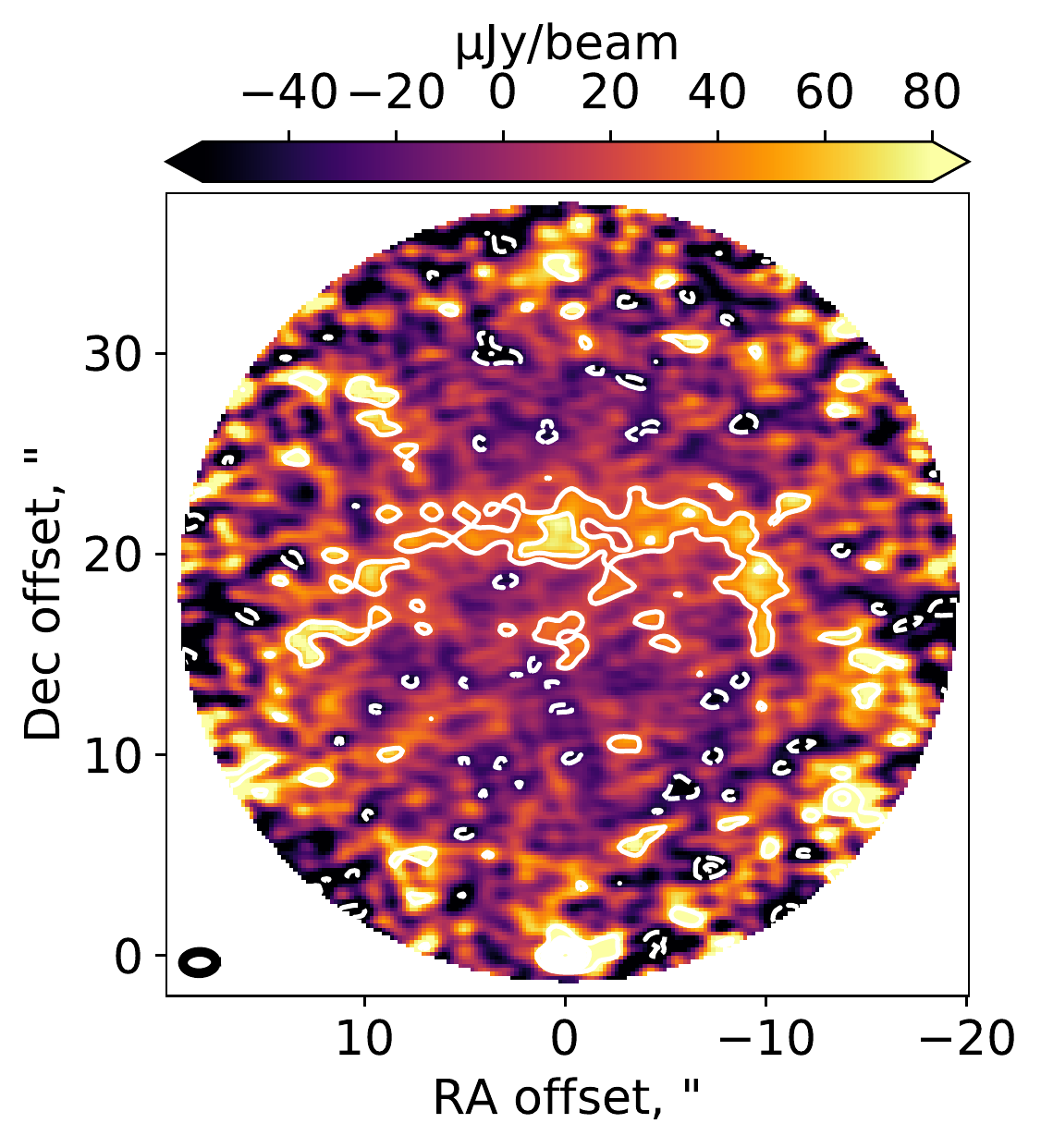}
	\includegraphics[width=0.33\textwidth]{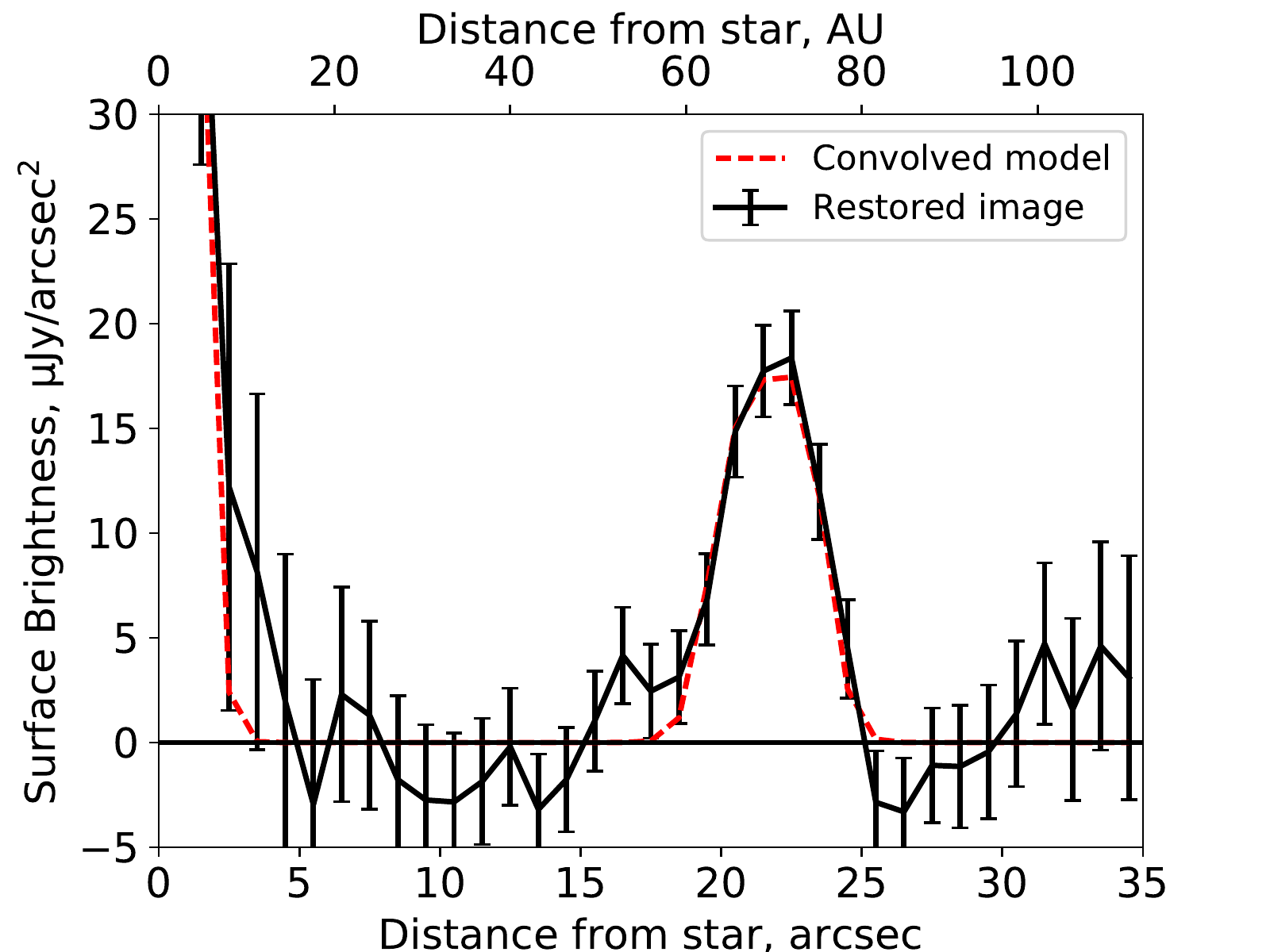}
	\includegraphics[width=0.33\textwidth]{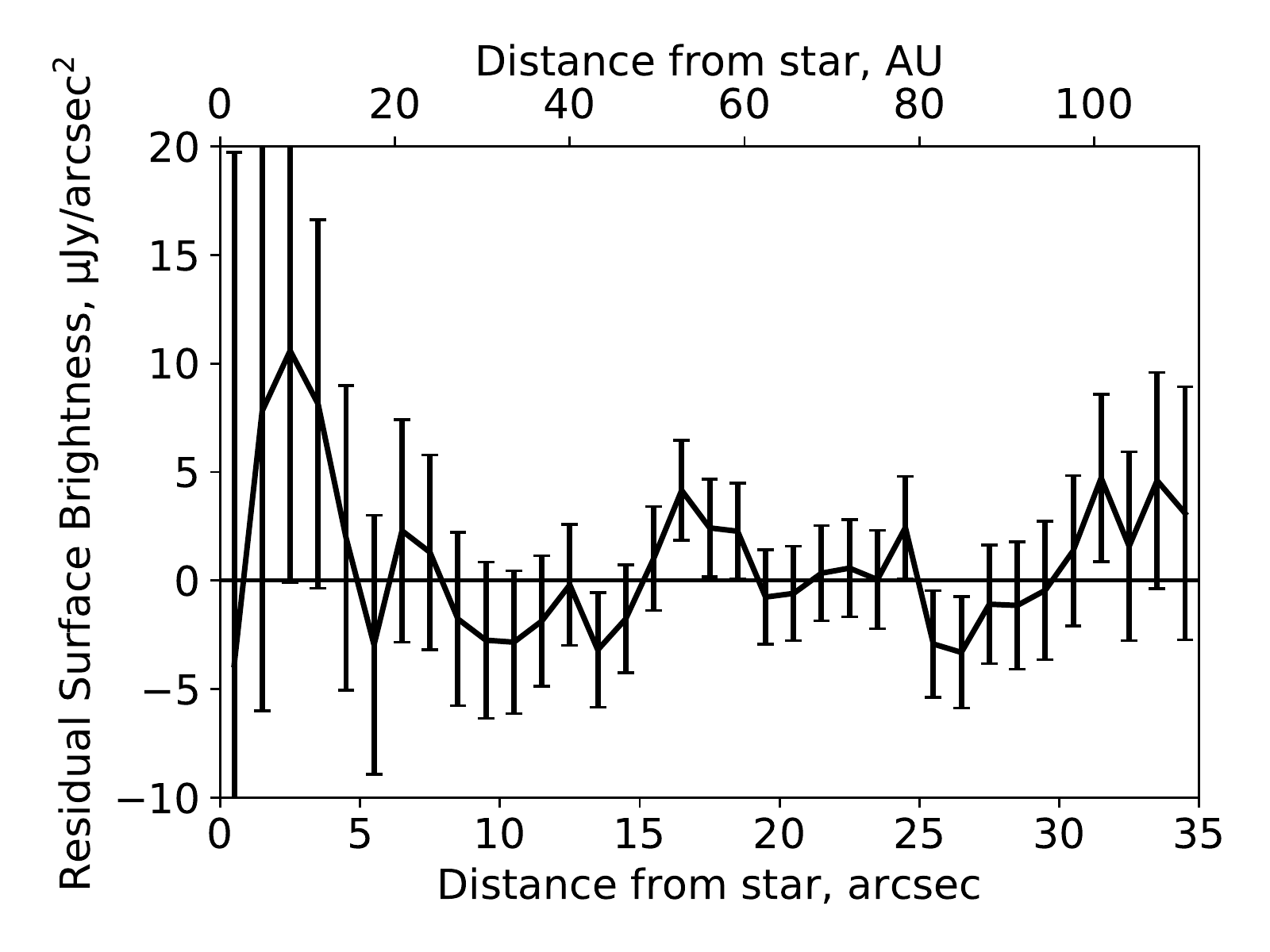} \\
	\includegraphics[width=0.33\textwidth]{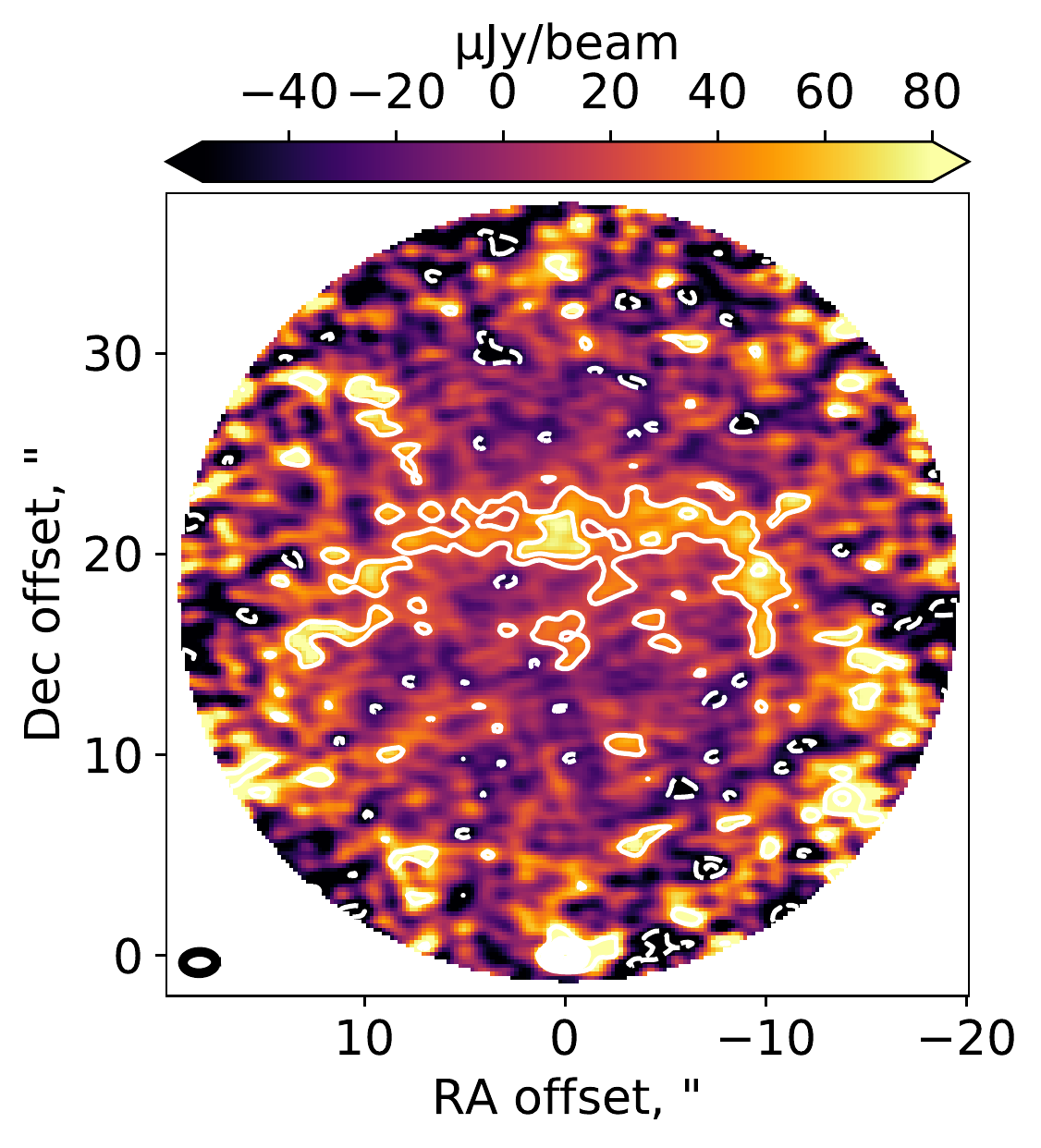}
	\includegraphics[width=0.33\textwidth]{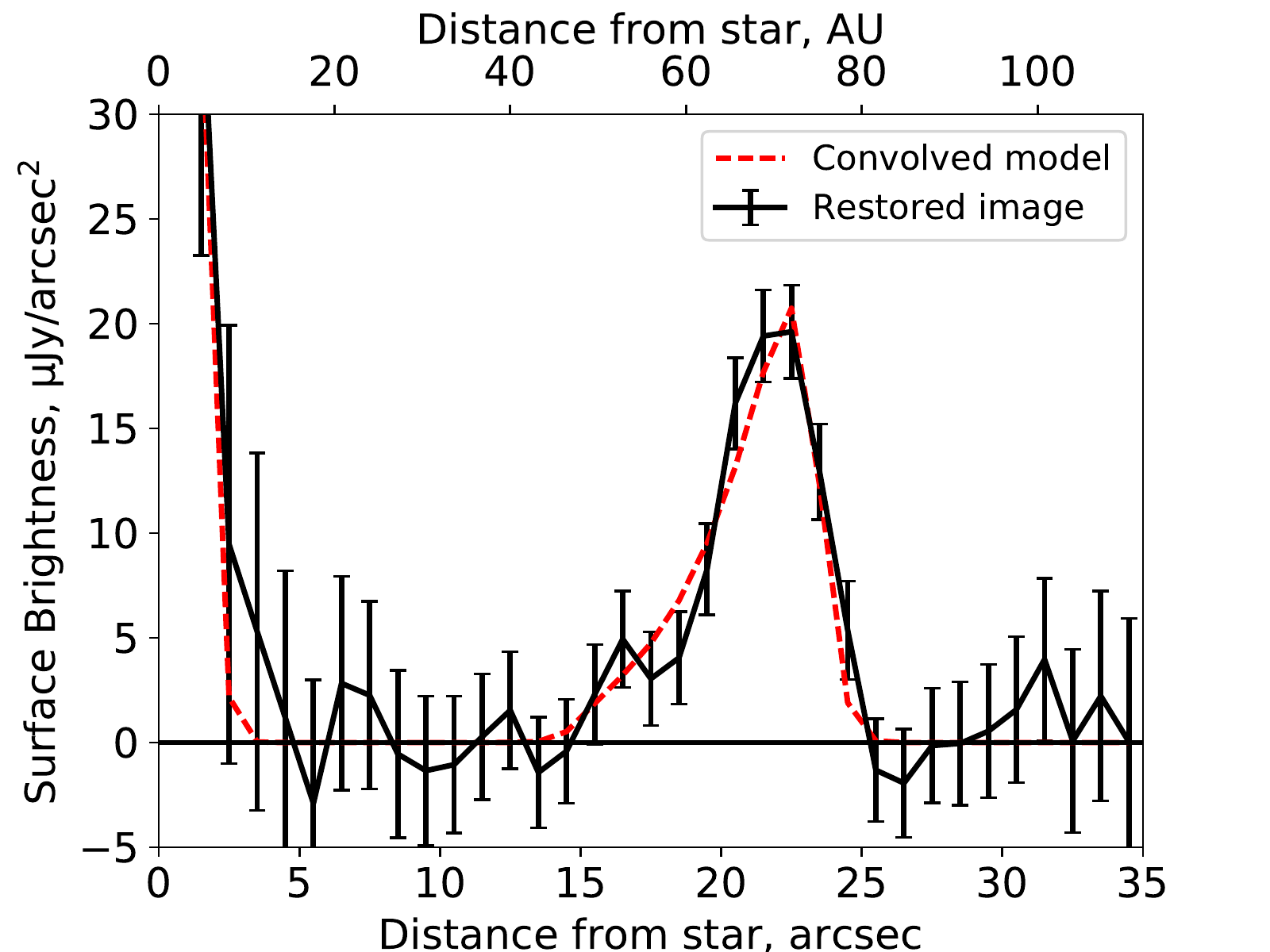}
	\includegraphics[width=0.33\textwidth]{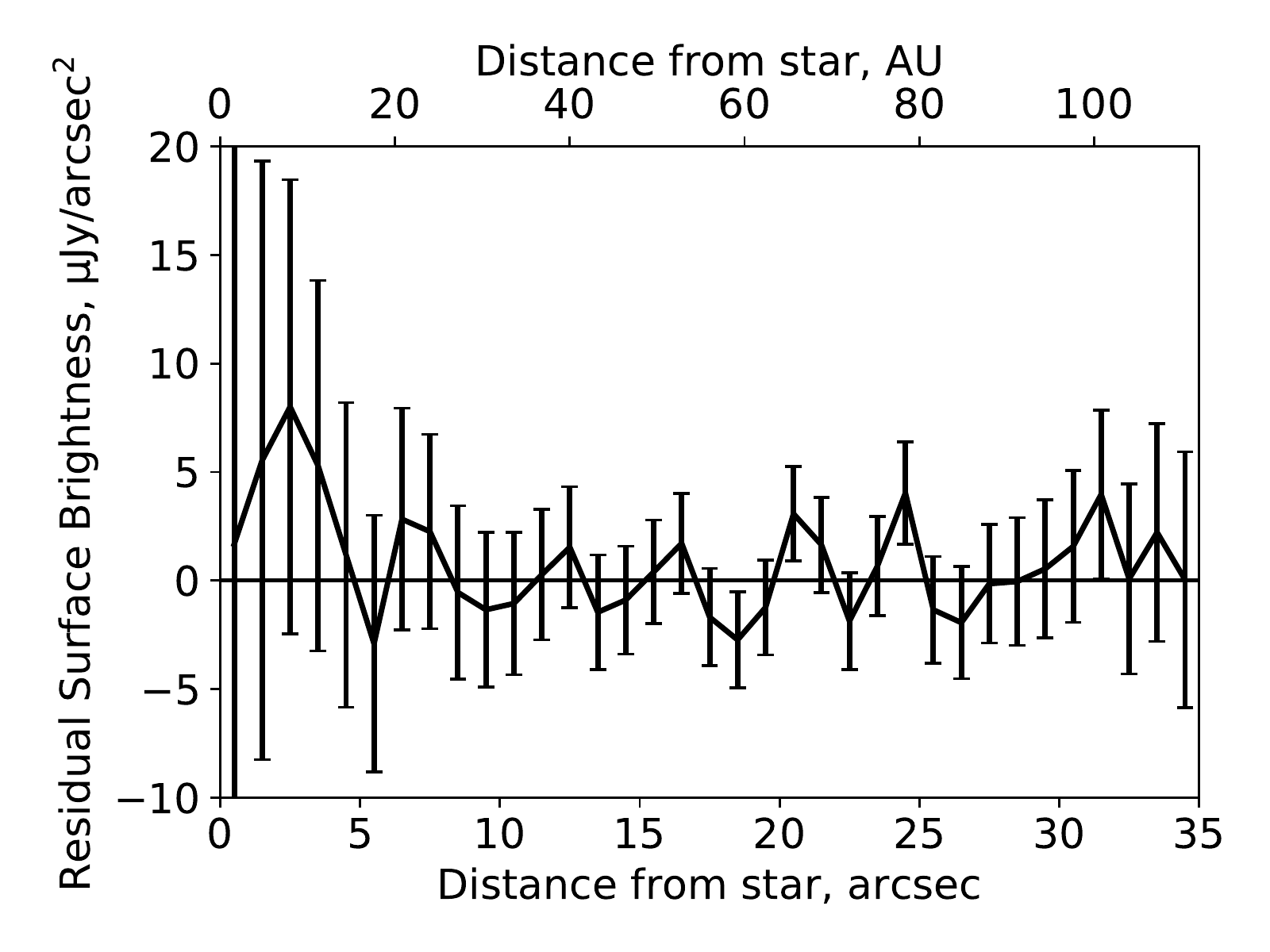} \\
	\includegraphics[width=0.33\textwidth]{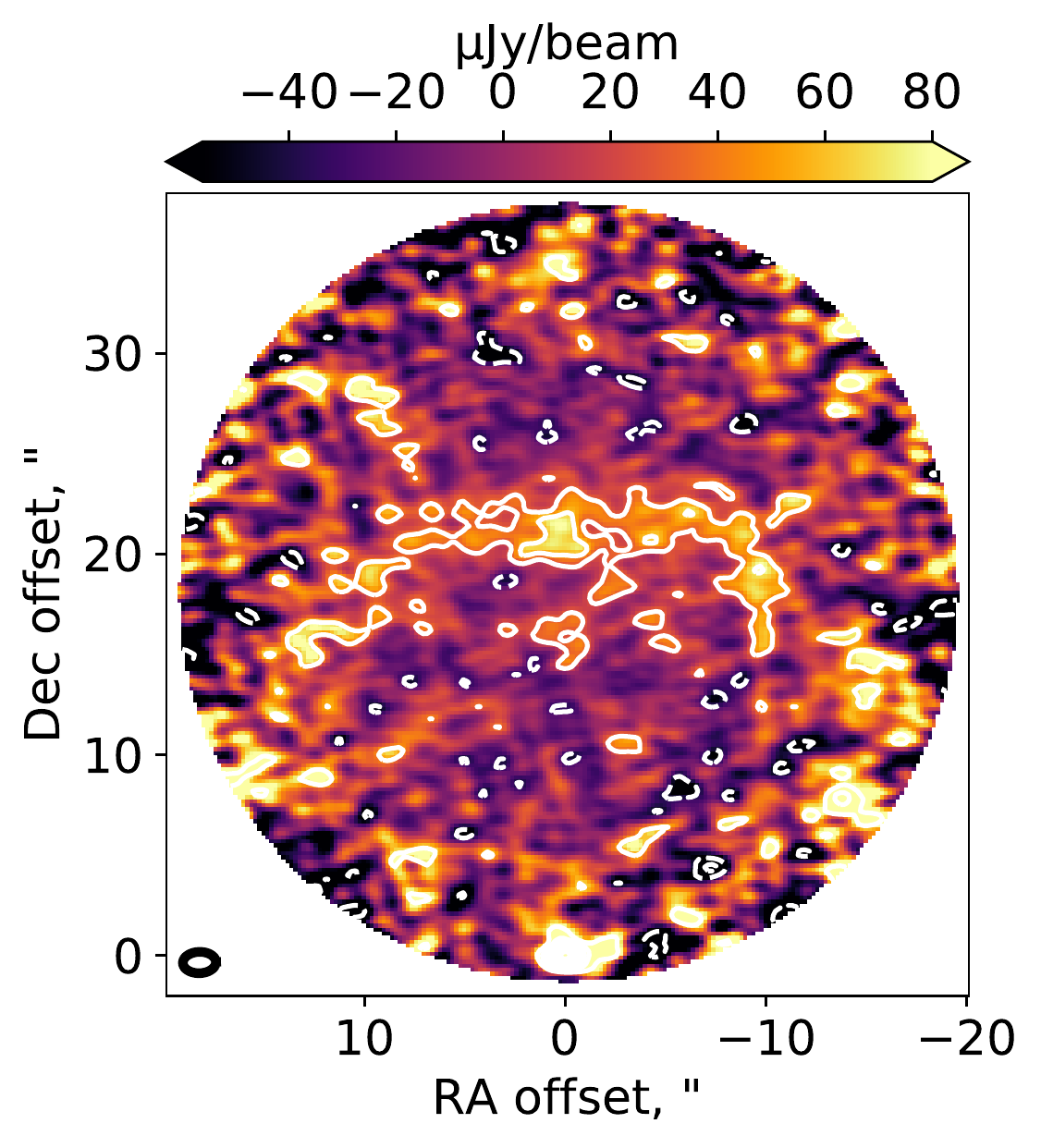}
	\includegraphics[width=0.33\textwidth]{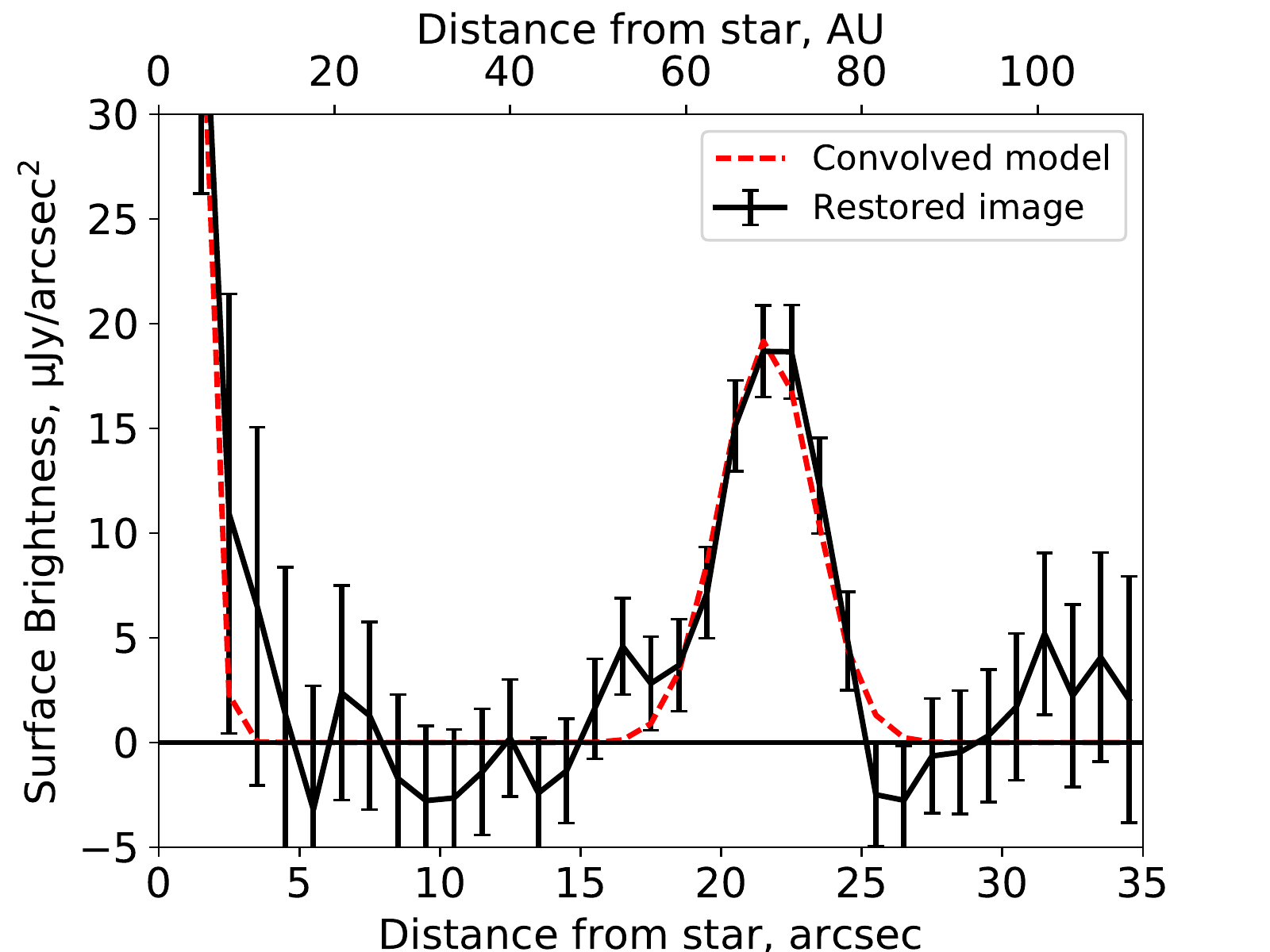}
	\includegraphics[width=0.33\textwidth]{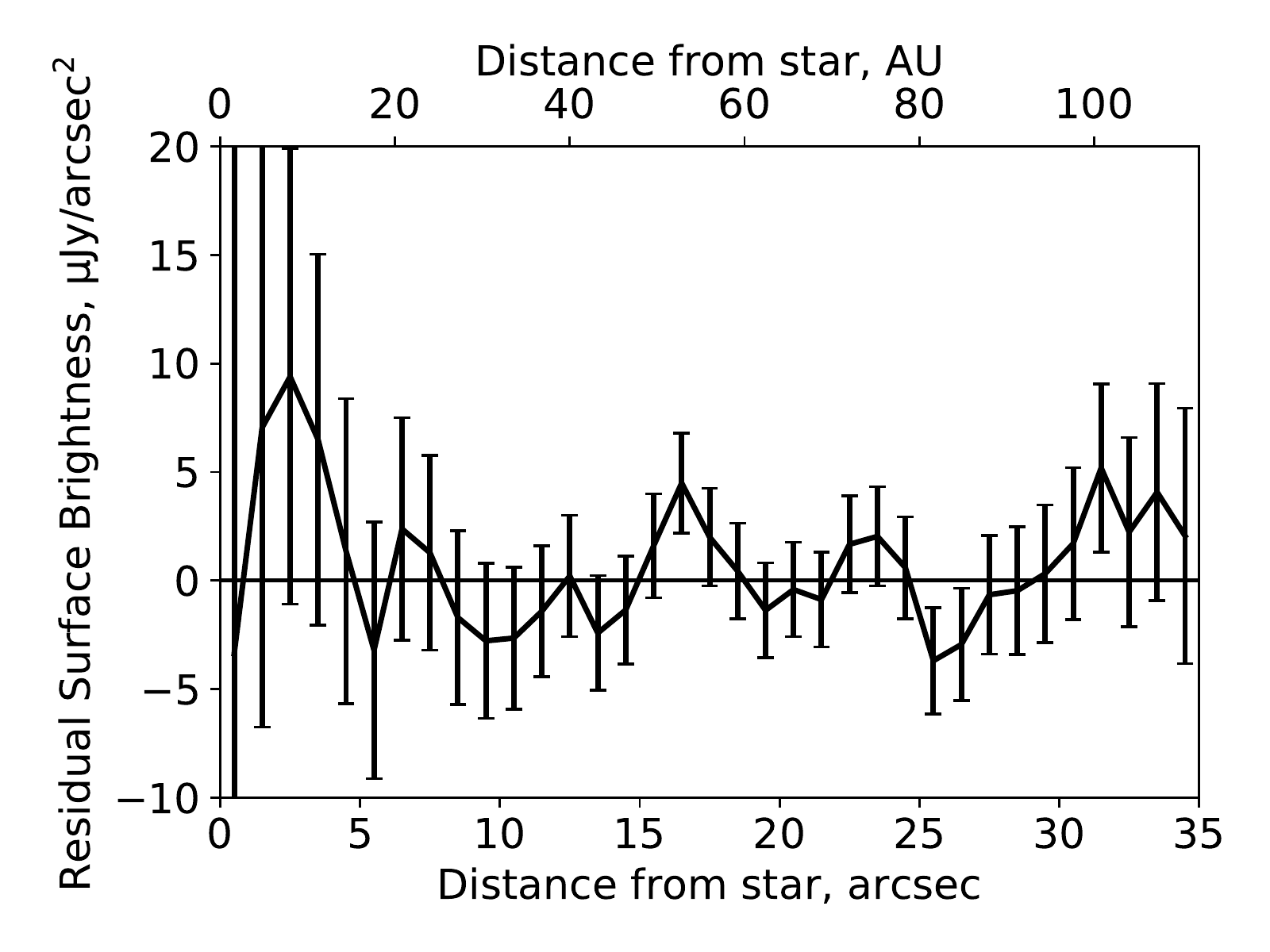}
	\caption{\emph{Left:} Primary beam corrected, restored image (i.e. best-fit model convolved with the synthesised beam and added to the residuals). Contours are in increments of $\pm2\sigma$. \emph{Middle:} Surface brightness distribution of the deprojected, restored image (solid line) and the model convolved with the synthesised beam (dashed line - red in the colour version). Error bars represent the noise (as calculated in section \ref{smod}) and do not take into account systematic uncertainties. \emph{Right:} Surface brightness distribution of the deprojected residuals. \emph{Top row:} Run A. \emph{Middle row:} Run B. \emph{Bottom row:} Run C.}
	\label{favg}
\end{figure*}

Given the low signal-to-noise ratio of our data, it is also useful to consider the radially averaged profiles. Comparing directly with the observations (figure \ref{fclean} left) is not very useful as this is a poor representation of the real radial profile due to the low level positive and negative structures in the PSF (figure \ref{fclean} middle) that result from the interferometric nature of the observations. 
We can reduce these interferometric artefacts by creating a restored image for which the PSF is the synthesised beam. This is created using our best fit results (see left-hand plots in figure \ref{favg}) in a similar manner to how the CLEAN image shown in figure \ref{fclean} is a restored image based on the CLEAN model. Since the interferometric artefacts cannot be removed completely without knowing exactly the real flux distribution, each restored image will be slightly different. We then deproject these images based on the best fitting $I$ and $\Omega$ parameters for each model and calculate the weighted average in each annulus (where annuli of 1\arcsec{} have been used here to provide the best compromise between signal to noise ratio and resolution). As the pixels of our image are smaller than the beam size, the noise in the final image is correlated and so we account for this by adjusting the uncertainty for each weighted average by a factor of $S_{ncr}$. As with the likelihood calculation, we only use pixels at a primary beam level of 20\% or higher in this calculation.

The middle column of figure \ref{favg} shows the deprojected radial profile for each restored image compared to the model and the right column shows the deprojected radial profile of the residuals. From these it is clear that the main ring has a steep outer edge, although both our sharp outer edge and Gaussian models reproduce it well. The inner edge is also reproduced well by both a sharp or Gaussian inner edge up to $\sim$19\arcsec. Interior to this the flux then rises and falls again, possibly suggesting an extra component just interior to the main ring, which we shall investigate further in the next section. However, the significance of this is only 2$\sigma$ and this interior emission and the inner edge are also consistent with a $\tau\propto R^7$ slope as shown by model B. This also shows why model B has a slightly lower $\chi_{red}^2$ than models A and C.

\begin{figure}
	\centering
	\includegraphics[width=0.23\textwidth]{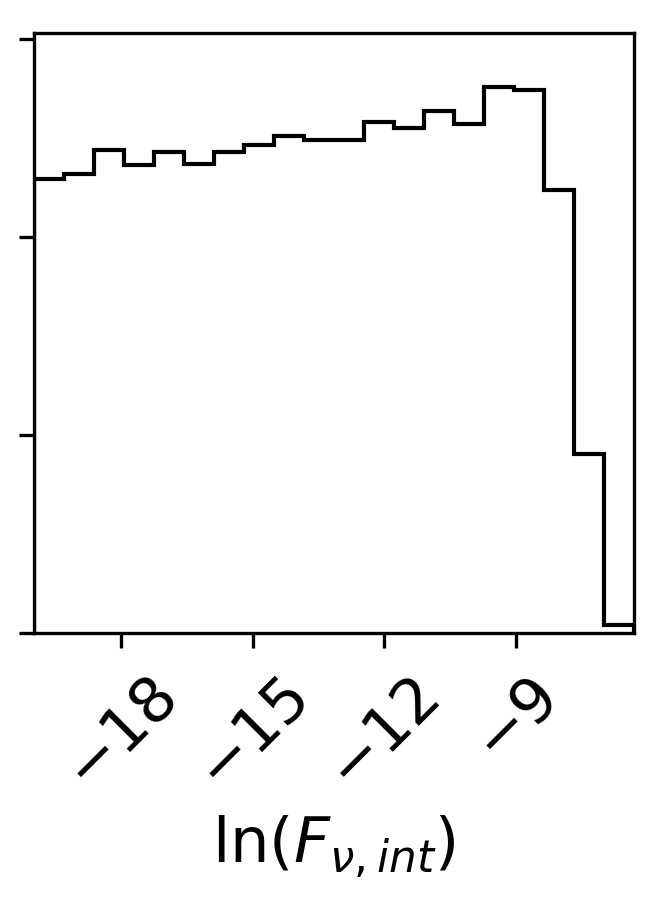} 
	\includegraphics[width=0.23\textwidth]{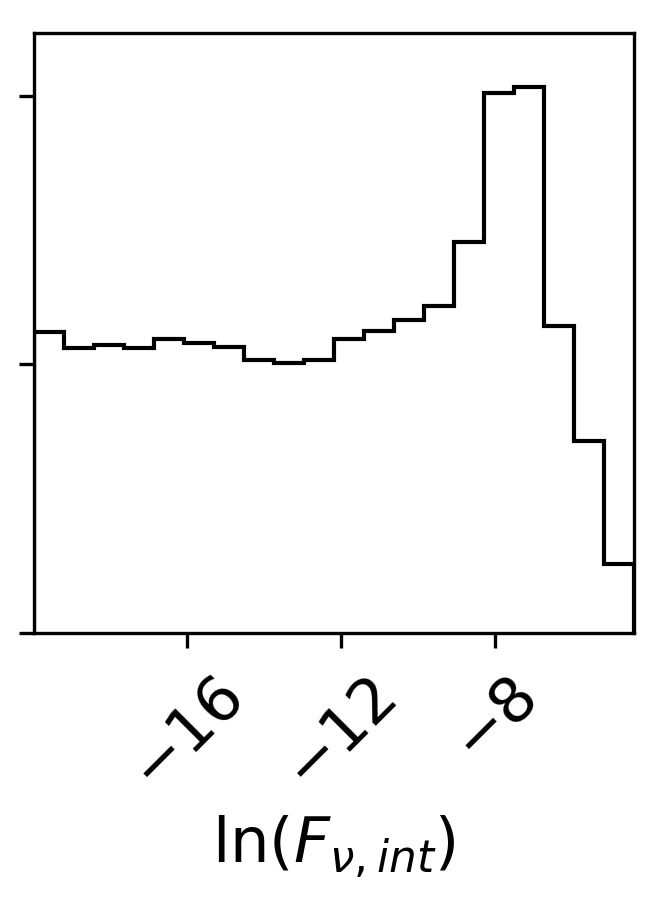}
	\caption{Posterior probability distributions for the interior belt flux density (in Jy) for run D (left) and run E (right). In both cases the flux density of the interior belt is consistent with 0, although in the latter case a peak is seen around 0.5~mJy.}
	\label{fprob}
\end{figure}

\subsection{Multiple components}
\label{smulti}
Although we cannot say for certain from these observations whether there are any further components between the main ring and the star, given that other observations by Spitzer \citep{backman09}, Herschel \citep{greaves14a}, LMT \citep{chavez16} and SOFIA \citep{su17} have detected such emission, it is worthwhile to use our data to determine any upper limits on such emission. Specifically, we wish to check two of the models suggested to explain the LMT observations. \citet{chavez16} suggest that the excess they see interior to the main ring could be fit by either a narrow ($\sim$10~AU) ring at around 20~AU or a wide distribution with a $R^{-3.5}$ radial profile and an inner edge around 14~AU. To test these we assume that the main belt is described as in the best fit for run A and that the interior dust has the same inclination and position angle as the main belt. The geometric parameters of the interior component are allowed to vary slightly from those in \citet{chavez16} with the first model (run D) having prior limits of $10<R_{in, int}<30$ and $R_{in, int}<R_{out, int}<30$ and the second model (run E) having prior limits of $0<R_{in, int}<20$, $40<R_{out, int}<75$ and $-5<\gamma_{int}<-2$.

\citet{chavez16} find the total flux density of an interior component (at 1.1~mm) to be 1.7-3.3~mJy, depending on the contribution from the stellar chromosphere (see section \ref{sstar}). Extrapolating this flux density to 1.3~mm using a spectral slope of $\lambda^{-3}$ (based on their spectral slope fit) means we expect $F_{1.3mm}=$1-2~mJy. For our data, in both cases these extra components have a total flux density that is consistent with 0~mJy (see figure \ref{fprob}). Run D seems especially unlikely as a narrow belt should be relatively easy to detect in our observations and models with a total flux density of this component $>$0.8~mJy are ruled out at the 99.7\% level. If a narrow belt does exist at this distance, then the spectral slope must be steeper than the $\lambda^{-3}$ fit of \citet{chavez16} and they must have underestimated the contribution of the stellar chromosphere. A wider disc as in run E can more easily be missed by interferometer observations and so the limits on this are less constraining with only total flux densities $>$8.5~mJy being ruled out at the 99.7\% level and a peak in the posterior probability distribution around 0.5~mJy.

\begin{figure*}
	\centering
	\includegraphics[width=0.32\textwidth]{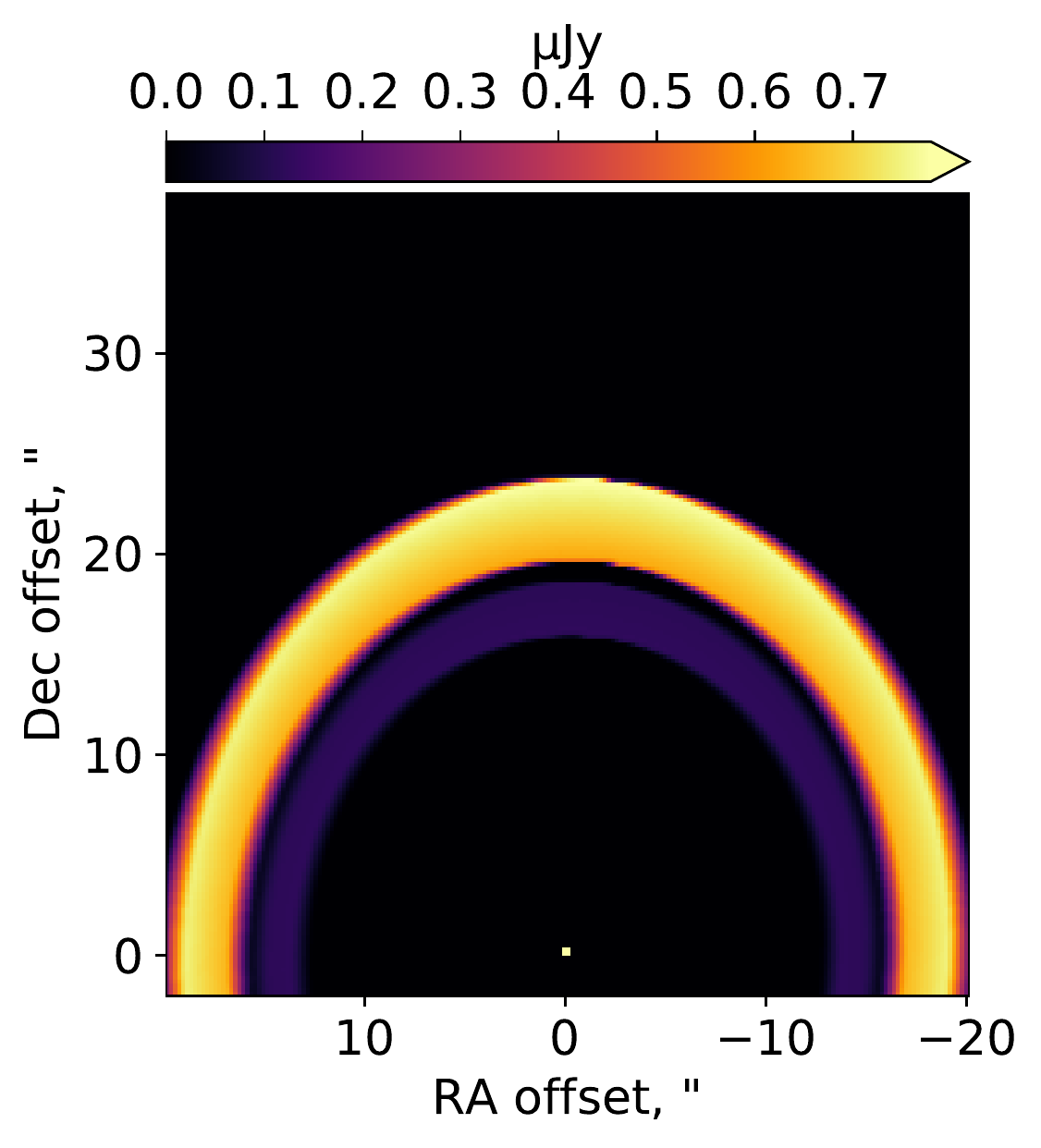}
	\includegraphics[width=0.32\textwidth]{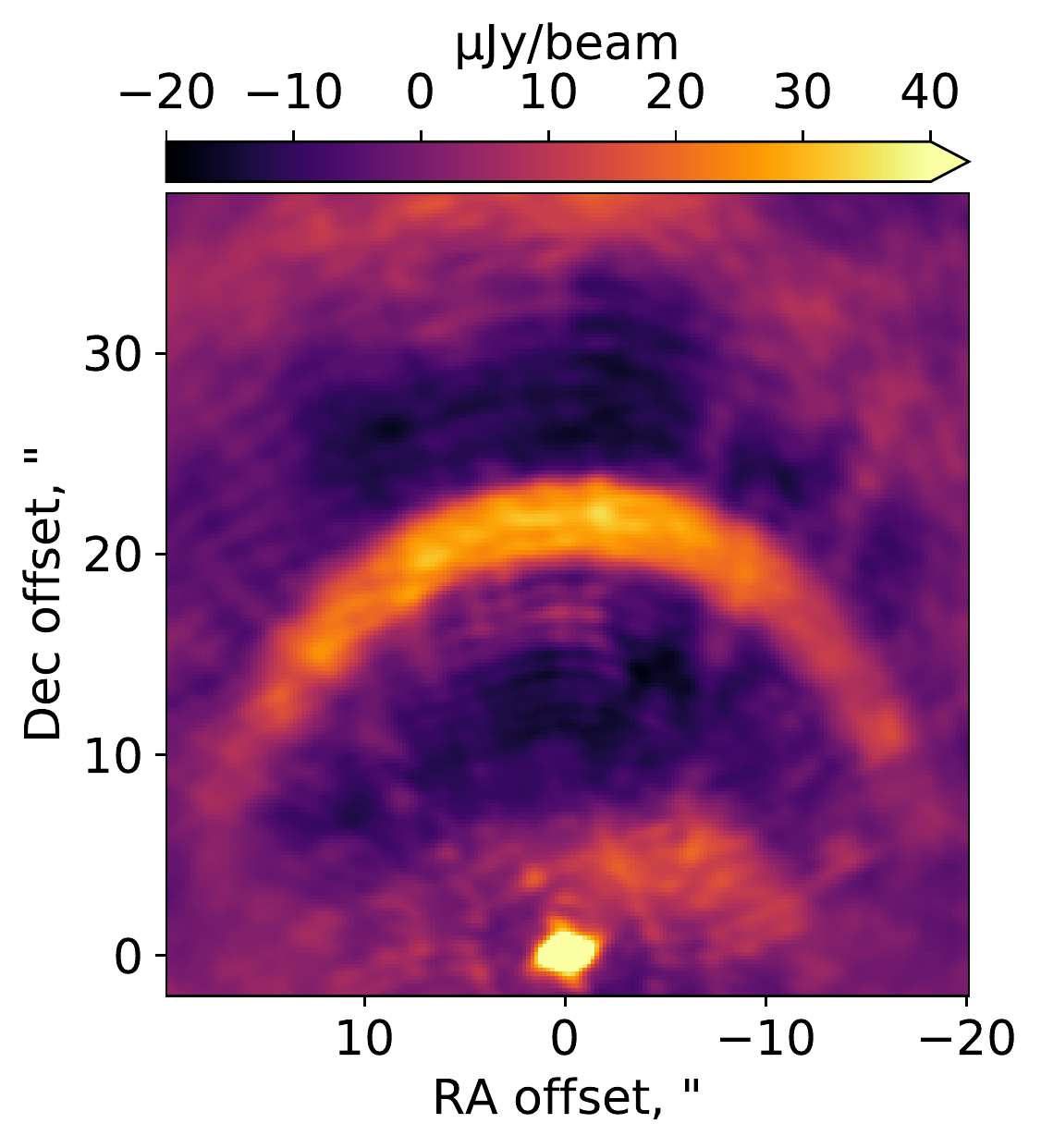}
	\includegraphics[width=0.32\textwidth]{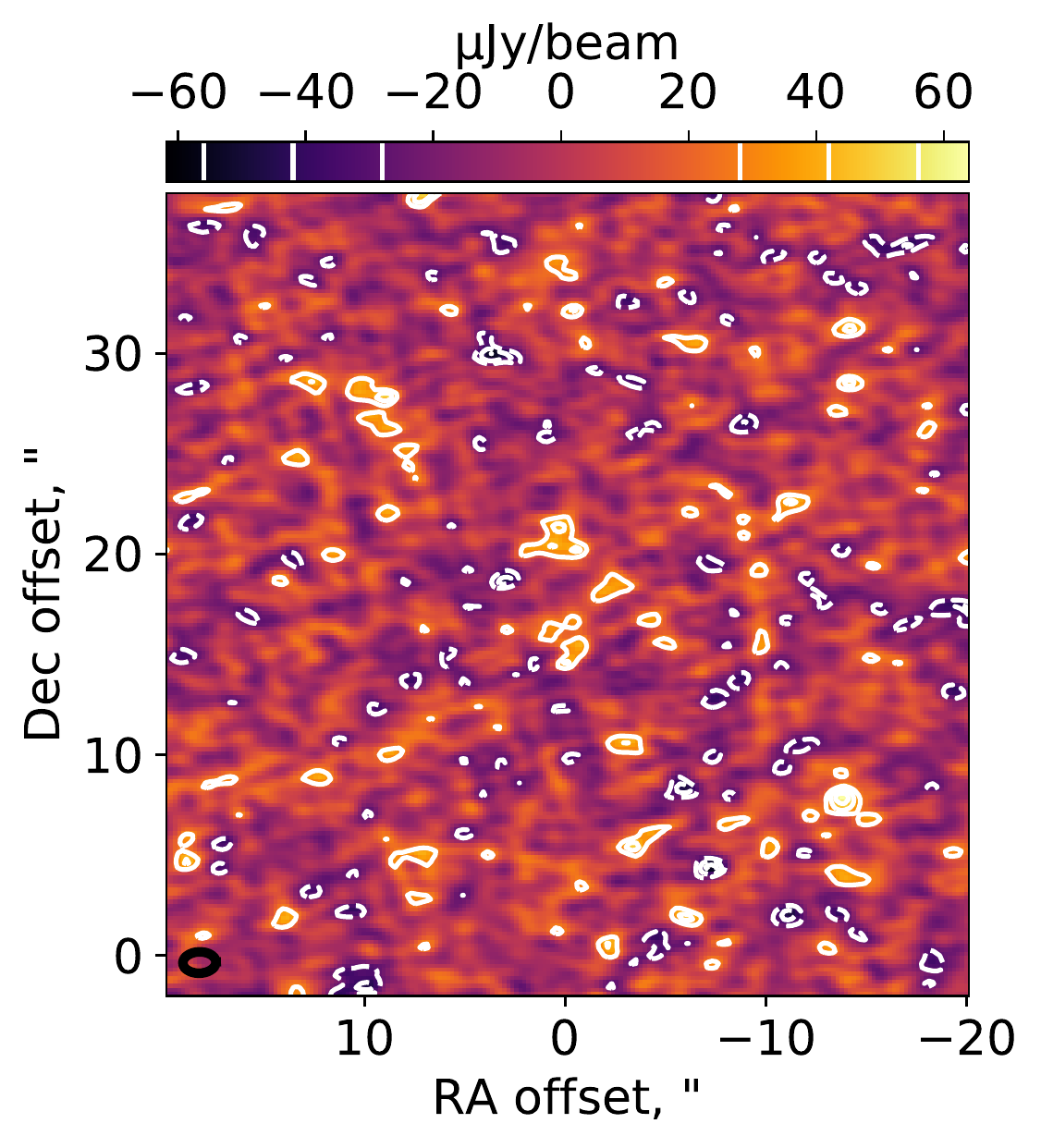} \\
	\includegraphics[width=0.32\textwidth]{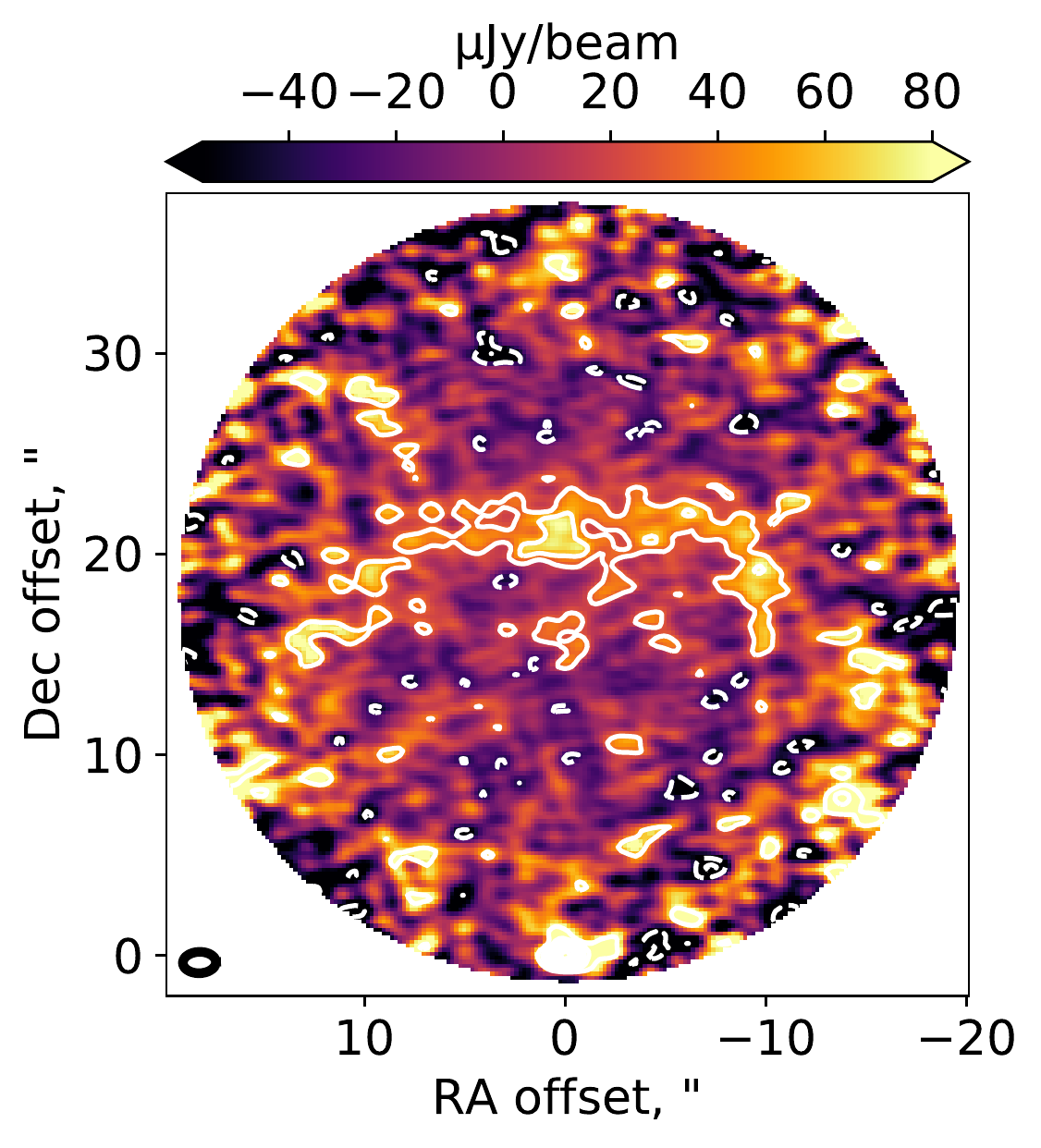}
	\includegraphics[width=0.33\textwidth]{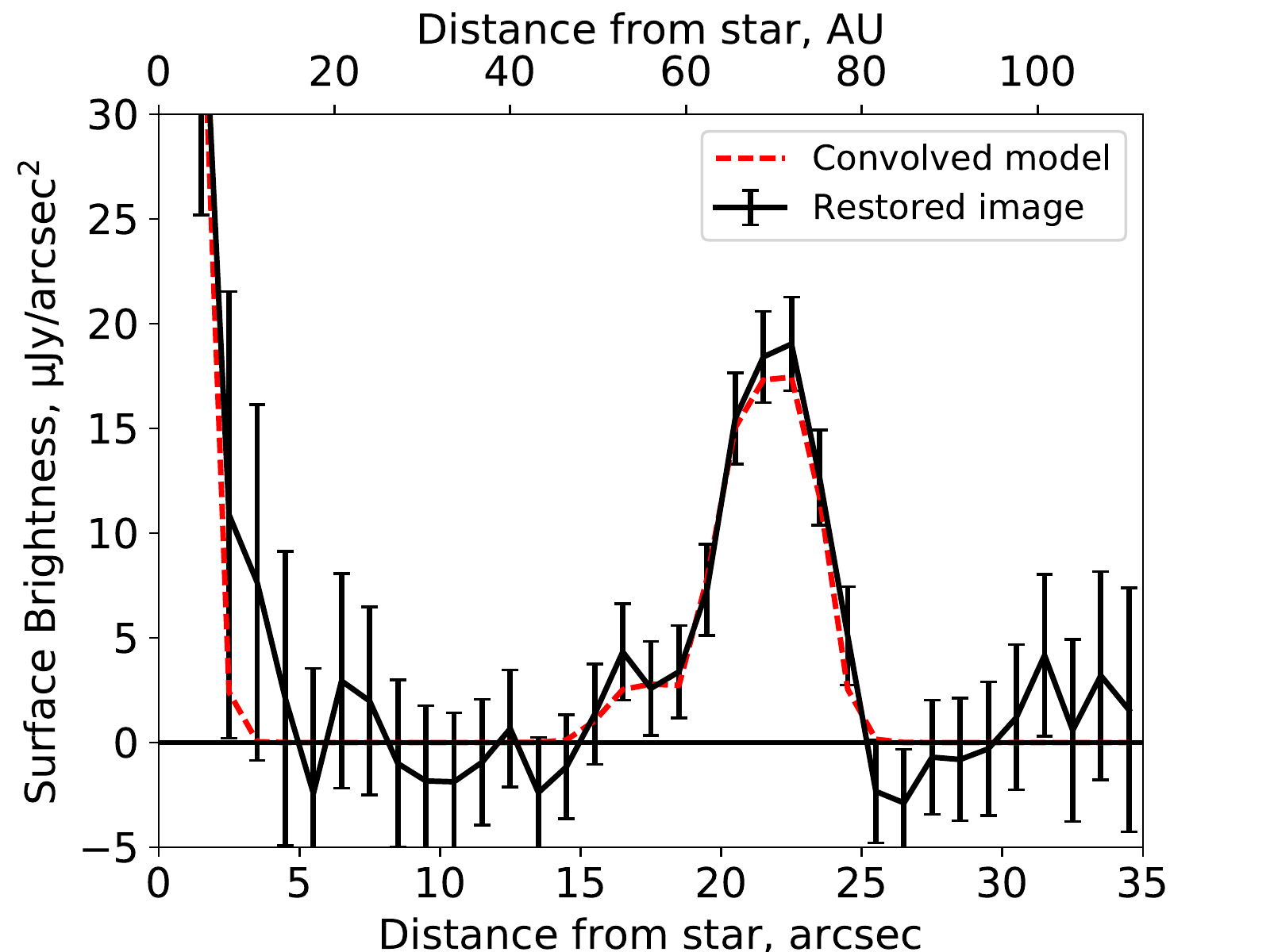}
	\includegraphics[width=0.33\textwidth]{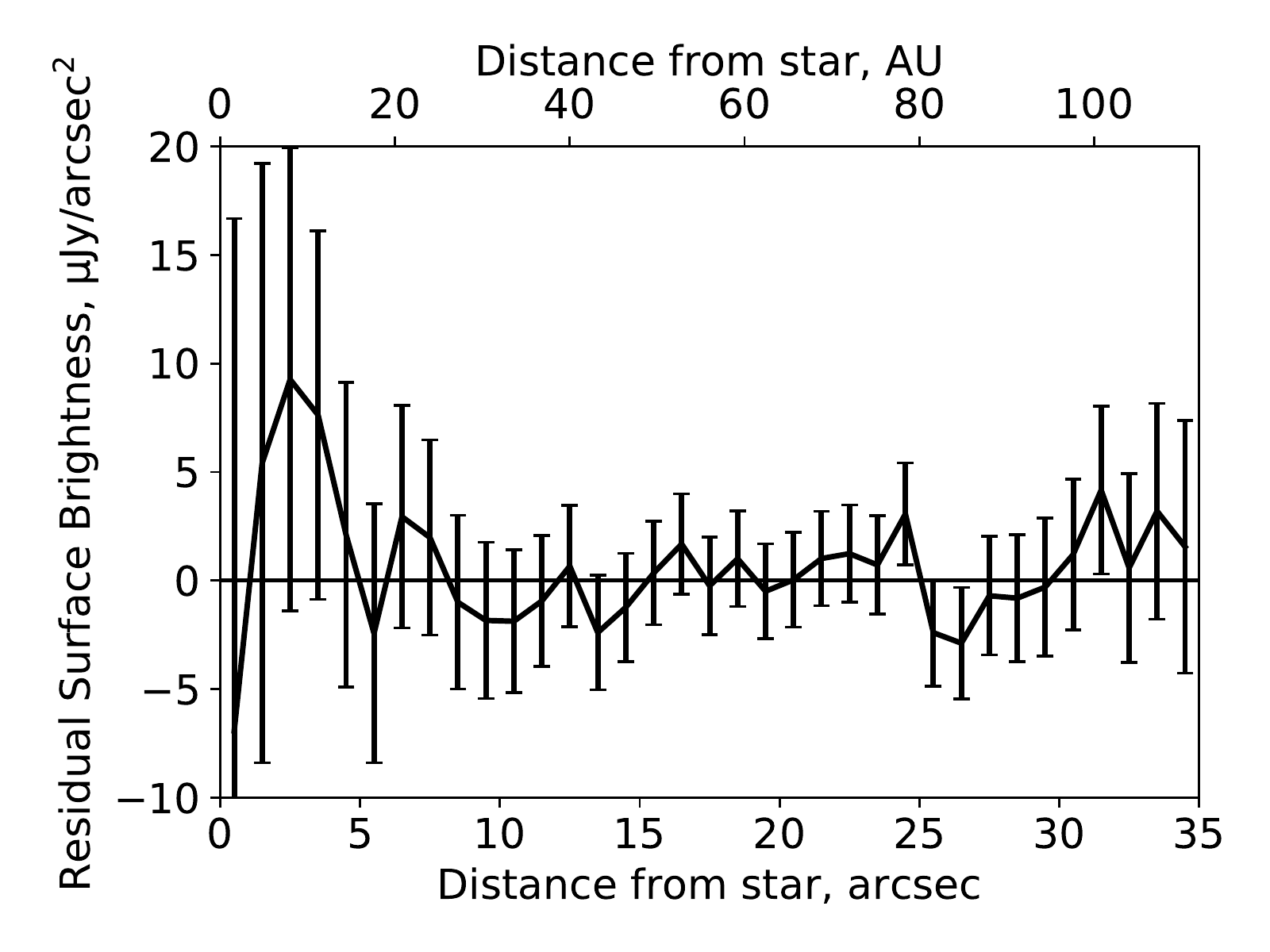} \\
	\caption{These plots show the best-fit model for run F, which is a fit using two narrow rings. \emph{Top Left:} Model. \emph{Top Middle:} Model convolved with the dirty beam and attenuated by the primary beam. \emph{Top Right:} Residuals after the best-fit model is subtracted from the image. Contours start at $\pm2\sigma$ and increase in increments of $1\sigma$.\emph{Bottom Left:} Primary beam corrected, restored image. Contours are in increments of $\pm2\sigma$. \emph{Bottom Middle:} Surface brightness distribution of the deprojected, restored image (solid line) and the model convolved with the synthesised beam (dashed line). \emph{Bottom Right:} Surface brightness distribution of the deprojected residuals.}
	\label{ftworing}
\end{figure*}

From the plots in figure \ref{favg}, there is a tentative sign of emission just interior to the main belt with a peak of 2$\sigma$ at 17\arcsec{}. Given the uncertainties, this could be explained by a rising power law as in run B. Alternatively, this could be a sign of a faint belt just interior to the main belt. We, therefore, conduct one further run (F) that is similar to run D, except that the priors force it to be further from the star: $35<R_{in, int}<70$, $R_{in, int}<R_{out, int}<70$ and $-10< \ln F_{\nu}<1$. This results in a final fit where $R_{in, int} = 51^{+3}_{-4}$~AU, $R_{out, int} = 59^{+5}_{-5}$~AU and $F_{\nu, int} = 0.7^{+0.4}_{-0.4}$~mJy. The plots for this best fit of this run are shown in figure \ref{ftworing}. Whilst this model does a better job of fitting the radial profile than a single component alone, it offers just a very slight improvement on the single component models with a $\chi^2_{red}=1.086$ and so further observations are necessary to determine the dust distribution at this inner edge.

As noted in section \ref{ssingle}, there are three prominent residuals seen after finding a best fit model. Given their prominence, it is possible that they may be affecting the conclusions drawn about the radial profile of the ring. S3 is found to have a negligible influence on the radial profile, likely because it is a point source. S1 and S2, however, have a stronger influence. If we consider just a 10\degr{} sector north of the star then there are clear peaks at 16\arcsec{} and 21\arcsec{}. If we consider the azimuthal average not including these central 10\degr{}, then the flux just interior to the main ring is diminished. Of course, removing the part of the image with the highest sensitivity, also means the uncertainties then increase and the best fit for run F is still consistent within the 1$\sigma$ uncertainties. Observations of the rest of the ring at a constant sensitivity are necessary to determine whether the emission just interior to the main ring is only present to the north of the star or extends all around the star. 

\section{CO mass limit}
CO has not previously been detected in the \epseri{} system \citep{yamashita93,coulson04} and, as noted in section \ref{sobs}, there are no obvious signs of CO in the data cube produced during the reduction process for our data. This is unsurprising as very few debris disc systems have shown signs of gas and those that have are typically young ($\lesssim$40~Myr), A star systems \citep[e.g.][]{lieman16}. Nonetheless, CO gas has been tentatively detected around a couple of F star systems observed with ALMA by averaging around the disc \citep{marino16,marino17} and so we follow a similar method here. 

First, we need to make some assumptions about how the gas is distributed. Given the age of the system, we will assume that any gas is second generation, released from exocometary ice through the collisional cascade of planetesimals in the disc \citep{zuckerman12,matra15,kral16}. Therefore we assume the CO to be co-located with the dust of the best-fit model for run A, i.e. in a uniform ring between 63 and 76 AU. The channels containing the CO (2-1) line can be found by considering the barycentric radial velocity that the gas would have if it is moving at Keplerian velocity. The largest Keplerian velocity will be for gas at the inner edge and at the ansa. Here the Keplerian velocity is 1.9~km\,s$^{-1}$, although since we do not know in which direction the disc is rotating, this velocity could either be positive or negative. \epseri{} itself has a radial velocity of 16.43$\pm$0.09~km\,s$^{-1}$ \citep{nidever02} and so we consider the channels between 14.5 and 18.3~km\,s$^{-1}$ as potentially including the CO emission and spectrally integrate over these (three) channels. We then correct the image for the primary beam and spatially integrate over the ring, but to avoid including too much noise only consider pixels that are within the 50\% power level of the primary beam. Finally this sum is divided by the number of pixels per beam to account for correlated noise. The uncertainty on the individual pixels is given by the primary beam corrected RMS. Then the uncertainties of the pixels are added in quadrature and divided by the square root of the number of pixels per beam to give the uncertainty in the region considered. This result is then extrapolated to the full ring by multiplying by the ratio of the ring area to the total area of the pixels considered in the calculation giving a final result of 0.13$\pm$0.11~Jy\,km\,s$^{-1}$. We have, therefore, not detected any CO gas in the disc and can place a 3$\sigma$ upper limit on the CO (2-1) line flux density within the main belt of 0.33~Jy\,km\,s$^{-1}$. 

To determine an upper limit on the CO mass from this we follow the method of \citet{matra15}, considering both the case of low and high gas density. If the gas density is low then the excitation of the gas will be dominated by radiation rather than collisions and the CMB will be the dominant radiation source for low-energy rotational transitions at these wavelengths. Under these assumptions, the upper limit on the line flux density given above corresponds to a 3$\sigma$ CO mass upper limit of 9.6$\times10^{-7}$~M$_\oplus$. If, instead, the gas density is high, then the CO excitation is collision dominated and the gas can be treated as in local thermodynamic equilibrium (LTE). The mass is then dependent on the kinetic temperature of the gas; the minimum mass for our flux upper limit is achieved for a temperature of 16.6 K, corresponding to the energy of the upper level (J=2) of the transition. These assumptions then give a 3$\sigma$ CO mass upper limit of 2.6$\times10^{-8}$~M$_\oplus$. In other words, the radiation dominated regime gives a conservative upper limit, but in a secondary gas scenario we expect CO to be released with other species that will contribute to its collisional excitation \citep[as is the case for $\beta$ Pic,][]{matra17}, meaning that this upper limit is likely closer to the LTE value. 

By assuming that the bodies that make up the collisional cascade contain a similar fraction of CO as those in the Solar System and comparing the production rate of CO from a collisional cascade to the photodissociation timescale, \citet{kral17} show that \epseri{} should have roughly 10$^{-9}$~M$_\oplus$ of CO. Our non-detection is, therefore, consistent with the steady-state production of CO.

\section{Discussion}
\subsection{The cold ring}
The high resolution and sensitivity of our observations gives us greater precision of the parameters of the cold ring than previously available. Here we will discuss how these compare to previous observations and what further information we can learn from them.

\subsubsection{Radial profile}
\label{srad}
We find that the main belt is centred at 69~AU with a width of between 11 and 13~AU depending on the model. By considering the radial profiles in figure \ref{favg}, it is clear that both the inner and outer edges are steep. At the inner edge, however, there are signs of more emission suggesting either a shallower slope or a separate component just inside the main belt with an unresolved gap between the two. Whilst our outer edge seems to be in agreement with previous observations, other observations have found the inner edge and peak to be somewhat closer in than found here with values between 35 and 65 AU for the inner edge and 57 and 64~AU for the peak \citep{greaves98,greaves05,backman09,lestrade15,macgregor15}. This discrepancy could be because the lower resolution of previous observations made it impossible to distinguish the main belt, which is only around 4\arcsec{} wide, from the emission just inside it, strengthening the likelihood that the tentative detection of dust inside of the main belt is real. 

Previous observations \citep{backman09,greaves14a,chavez16,su17} have found evidence for emission interior to 15\arcsec{}. We do not find evidence for this in our data meaning that the dust at these distances from the star must be spread out in a wide disc and/or faint at this wavelength. 

\begin{figure}
	\centering
	\includegraphics[width=0.48\textwidth]{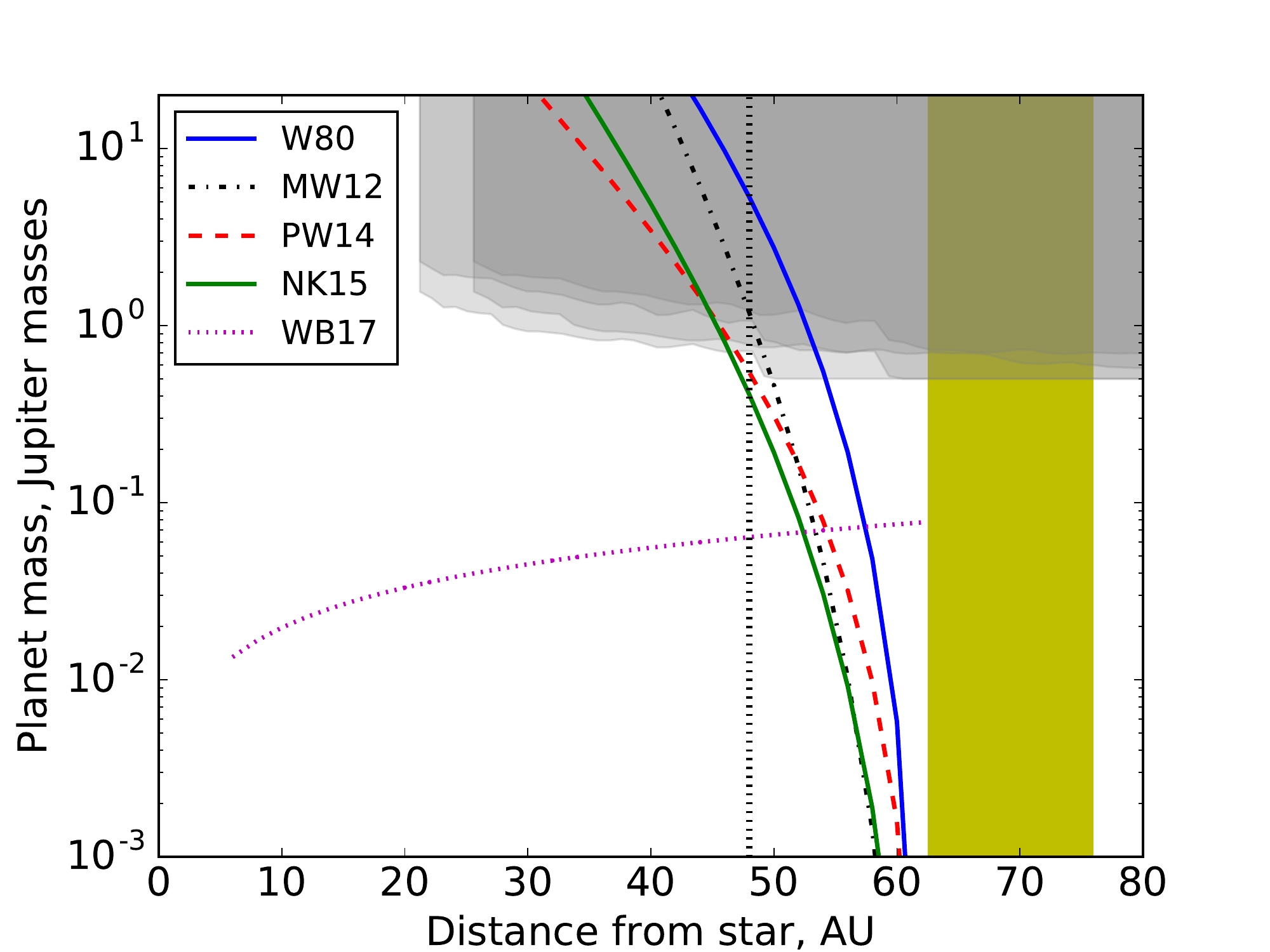}
	\caption{Dynamical estimates for the mass of a planet as a function of semi-major axis compared to observational limits. The lines given are based on the equations of \citet[W80;][]{wisdom80}, \citet[MW12;][]{mustill12}, \citet[PW14;][]{pearce14}, \citet[NK15;][]{nesvold15} and \citet[WB17;][]{wyatt17}. The vertical (yellow) shaded region represents the location of the main ring and the vertical dotted line at 48~AU shows where a planet would need to be to have 3:2 and 2:1 mean motion resonances that coincide with the discs inner and outer edges. The observational limits (represented by the grey shading) are from \citet{janson15}. }
	\label{fplloc}
\end{figure}

Such a large clearing interior to a narrow ring strongly implies the presence of planets that have cleared the region interior to the belt. Through scattering, planets clear the chaotic zone around them out to $\delta a$. \citet{wisdom80} analytically derived this to be
\begin{equation}
 \delta a=a_{pl}C\mu^{2/7}
\label{ewis}
\end{equation}
where $a_{pl}$ is the semi-major axis of the planet (in AU), $C$ is a coefficient found to be 1.3 by the approximate scaling theory of \citet{wisdom80} \citep[although other authors prefer values as high as 2; e.g.][]{quillen06,chiang09}, $\mu$ is the ratio of the planet mass to the stellar mass \citep[0.82~M$_\odot$ for \epseri;][]{baines12}. Assuming the inner edge is defined solely by the outermost planet in the system \citep[at least two planets are necessary to explain the extent of the clearing; ][]{deller05}, we use this to calculate the mass of the planet as a function of its semi-major axis as shown in figure \ref{fplloc}. The mass required for a given semi-major axis can be reduced by taking into account the eccentricity of the planetesimals \citep{mustill12}, the eccentricity of the planet \citep{pearce14} and collisions in the disc \citep{nesvold15}. Given the width of the ring, the planetesimals must have an eccentricity less than 0.1 and so we use this value in equation 10 of \citet{mustill12} to show the effect of a disc of eccentric planetesimals in figure \ref{fplloc}. We know that there cannot be a highly eccentric planet in the system otherwise an offset between the disc centre and the star would be seen but this offset is found to be less than 9~AU \citep{macgregor15} and so the eccentricity must be less than 9~AU/$a_{pl}$. We use this relation along with equations 9 and 10 of \citet{pearce14} to show the effect of an eccentric planet in figure \ref{fplloc}. These estimates are all purely dynamical. \citet{nesvold15} show that collisions are also important in determining the effect of a planet on the disc as collisional destruction of grains is enhanced in mean motion resonances, thus widening the gap. They find that the power law index for $\mu$ in equation \ref{ewis} is then dependent on the age of the system relative to the collisional timescale. \epseri{} is estimated to be between 400 and 800~Myr \citep{mamajek08} and so we assume an age of 600~Myr when demonstrating the effect of collisions in figure \ref{fplloc}.

These equations only provide us with a relationship between the mass of the planet and its semi-major axis. By considering the shape of the inner edge it is possible to further constrain these parameters as a low mass planet close to the belt results in a sharper edge than a high mass planet further from the belt \citep{chiang09,mustill12}. The effect of drag forces also needs to be taken into account as these may be strong in the case of \epseri{} as shown by \citet{reidemeister11}. Whilst the effect on large grains is limited, they find that it can be enough to make the slope of the inner edge shallower (A. Krivov, private communication), producing a rising profile as in run B. On the other hand, if there is an extra ring as in run F, then this would be a strong sign of clearing due to a planet either within the gap or through resonances. If the extra emission is not a complete ring but concentrated (see section \ref{smulti}) then this could be a sign of Trojans as demonstrated in the simulations of \citet{nesvold15}. Given this uncertainty in the inner edge, further observations are necessary before any accurate predictions can be yielded from such an analysis. 

One way to potentially determine the planet's semi-major axis is to consider its effect on both the inner and outer edge of the belt. From our results we find a fractional belt width of $\Delta R/R=0.17$, making this one of the narrowest discs known \citep{lestrade15}. For comparison, the classical Kuiper belt in the Solar System extends between semi-major axes of $\sim$40 and 48~AU \citep[e.g.][]{gladman01} giving it a fractional width of 0.18. It is notable that the inner and outer edges of the Kuiper belt correspond to the 3:2 and 2:1 mean motion resonances with Neptune, potentially a result of Neptune's migration \citep{hahn05,levison08}. If the belt we see around \epseri{} is also bounded by the 3:2 and 2:1 resonances of an unseen planet, then the inner and outer edges should correspond to the same planet semi-major axis. Using the parameters for the narrow belt fit, the inner edge would be at the 3:2 resonance of a planet at 47.8$\pm$1.2~AU and the outer edge would be at the 2:1 resonance of a planet at 47.8$\pm$1.0~AU, perfectly matching and suggesting the chance of a planet at this distance. However, it may just be coincidence and there are also a couple of caveats to consider. Firstly, the resonances effect the distribution of the planetesimals as a function of their semi-major axis. The planetesimals are likely to have a non-zero eccentricity and so the radial distribution will be wider than the semi-major axis distribution. Secondly, during the migration of Neptune that created these boundaries for the classical Kuiper belt, Neptune also scattered objects onto eccentric orbits creating the scattered disc extending much further out than 48~AU and resulting in a radial distribution of planetesimals that is broader than when only the classical Kuiper belt is considered. Nonetheless, the radial distribution of the Kuiper belt is found to have a FWHM that is only marginally broader than the locations of these resonances \citep{vitense10}. If there is a planet at 48~AU then we can use figure \ref{fplloc} to estimate its mass to be between 0.4 and 5 $M_J$.

Attempts to detect a planet in the outer parts of the system have been made using the direct imaging technique with NaCo on the Very Large Telescope, the Spitzer Space Telescope and Clio on the MMT \citep{marengo06,marengo09,janson07,janson08,janson15,heinze08}. Whilst none of these have been successful, they do provide useful upper limits on the magnitude of any planets in the system, which can be converted to mass limits given a planetary evolution model and an age. \citet{janson15} provide the most constraining limits on planets in the system, which are shown by the grey region in figure \ref{fplloc}. Some uncertainty is shown in these limits for two reasons. Firstly, if we assume the planet has the same inclination as the disc then the limits are more constraining when it is north or south of the star than east or west. Secondly, the limits are more constraining if a young age (0.4~Gyr) is assumed than if an older age (0.8~Gyr) is assumed as younger planets are brighter. Also note that the planet models used by \citet{janson15} to convert from luminosity limits to mass limits only work for masses $>0.5$~M$_J$ and so the limits for the 0.4~Gyr case beyond $\sim$50~AU could actually be more constraining than shown here. 

Combining the dynamical and observational constraints we see that if a planet is responsible for the location of the inner edge of the disc then it must be further out than $\sim$45~AU and have a mass $<1.3M_J$. If it is located at 48~AU, as suggested above, then these limits strongly rule out planets $>1.2M_J$ and potentially place limits down to $M_{pl}>0.7M_J$ if the system is only 400~Myr old in addition to the lower limit of 0.4~M$_J$ from the dynamical analyses noted above. It is important to note that all of the above only considers the effect of one planet and it is possible for other planets in the system to contribute to shaping the inner edge of a debris disc, thereby meaning that the outermost planet does not need to be so massive.

In section \ref{smulti} we discussed how there may be dust interior to the belt. If this is produced by planetesimals in that region then this suggests that planets have not completely removed comets on planet crossing orbits and we can use equation 2 of \citet{wyatt17} to derive an upper limit for the planet mass as a function of semi-major axis. This is also shown in figure \ref{fplloc} assuming a $t_\star=0.6$~Gyr.

\subsubsection{Orientation to the line of sight}
Despite only observing part of the ring, we are still able to precisely measure the orientation of the disc to the line of sight. Our models result in an inclination of around 34$\pm2$\degr{} and position angle of $-4\pm3$\degr{} East of North. Previous work has found a range of different inclinations. The Herschel \citep{greaves14a} and LMT \citep{chavez16} results most closely agree with ours, both finding an inclination of around 30\degr{} (with uncertainties of around 5\degr{} and 10\degr{} respectively). The SMA result shows the largest inclination difference to ours -- 17$\pm$14\degr{} \citep{macgregor15} -- although the large uncertainties mean it is still consistent with our result. All observations so far have consistently given position angles around 0\degr, with \citet{chavez16} also finding a slight rotation West of North with $\Omega=-7\pm10\degr{}$.

In general, planetary systems are found to be roughly coplanar \citep{kennedy13a,greaves14} although this is not always the case \citep[e.g.][]{winn10} and so it is worth considering how the orientation of the disc compares to that of the planet and the star. \epseri{} b was first discovered by \citet{hatzes00} using the radial velocity (RV) technique. RV alone cannot determine the orientation of a planet's orbit, but combining with astrometry can provide this information. \citet{benedict06} combine the RV data with astrometry from the Hubble Space Telescope to calculate an inclination of 30.1$\pm$3.8\degr{} and \citet{reffert11} make use of Hipparcos to calculate an inclination of 23$\pm$20\degr, both of which agree with our result. However, it is not enough to just compare the inclinations as the disc and planet may be inclined in different directions. For a system to be coplanar, the longitudes of ascending node must also match. \citet{benedict06} and \citet{reffert11} determine the longitude of ascending node to be 254$\pm$7\degr{} and 282$\pm$20\degr{} respectively, considerably different to our position angle meaning that the system is not coplanar. Although the difference may, alternatively, be because there are issues with the RV analysis, that would result in a different orbital fit \citep{anglada12}.

The inclination of the star has been determined spectroscopically to be 30$\pm$20\degr{} \citep{campbell85,saar97}. More recently, the Microvariability and Oscillations of STars telescope has observed the star but different analyses provide different results with \citet{croll06} finding 30$\pm$3\degr, \citet{frohlich07} finding 45$^{+11}_{-19}$\degr{} -- albeit with a possible alternative fit of 72\degr{}  -- and \citet{giguere16} finding 69.5$_{-7.6}^{+5.6}$. The analyses are unable to provide any information on the direction in which the stellar axis is tilted. It is therefore not yet possible to say for sure whether the disc, known planet and star are coplanar.

\subsubsection{Disc flux density}
From our modelling, we find the total flux density of the disc to be between 8 and 10~mJy with uncertainties around 1~mJy (depending on the assumed model). As the SMA observations are at the same wavelength we can compare directly with these. As the SMA is also an interferometer, \citet{macgregor15}, like us, provide a flux density found through model fitting rather than aperture photometry, which they find to be 17$\pm$5~mJy. Whilst lower than this, our result is still within 2$\sigma$. However, a full SED analysis of all the infrared and millimetre photometry \citep{chavez16} finds an SED fit that is consistent with the SMA result and not consistent with our result. 

It is possible that our modelling approach is responsible for this discrepancy. \citet{white16} fit ALMA data of Fomalhaut using two different approaches. First fitting in the image plane, as we have done in this paper, and second fitting in the visibility plane. They find that the best fit disc flux density from the first approach is lower than both the value found from the second approach and an extrapolation from single dish measurements. To make sure that a similar issue is not the reason for our discrepancy, we re-run model A, this time fitting to the visibilities, and find only a marginally ($<$1\%) higher disc flux density (see appendix \ref{avfit}). 

The actual reason for our lower total flux density is likely twofold. Firstly, we have only observed part of the ring and assumed there to be no large variations around the ring. \citet{greaves14a} and \citet{chavez16} both find the South-East of the disc to be brighter than the North. This could be due to an actual asymmetry in the disc that our model does not take into account, perhaps due to apocentre glow \citep{pan16}, or due to a background source. Observations of the rest of the ring are necessary to confirm whether this is the case. Secondly, the interferometer set-up used for these observations has a shortest baseline of 15~m meaning that we are not sensitive to emission on scales larger than 18\arcsec. Whilst our modelling procedure takes account of this so that the fitted flux density is accurate for the given model, we could easily be missing larger scale components that do not show up in our data and so we are unable to place any strong constraints on them. For instance, we showed in section \ref{smulti} that adding an interior component of up to a few milli-Janskies would still be consistent with the data. Observations with shorter baselines are necessary to determine whether we are missing flux. 

\subsection{Stellar flux}
\label{sstar}
As mentioned in section \ref{sobs}, the flux density at the star's location is found to be $F_{cen}=820\pm68\,\mu$Jy. The flux density of the stellar photosphere at 1.3~mm is expected to be $530\pm11\mu$Jy based on a Kurucz atmosphere model \citep{macgregor15}, showing that there is an excess over the photosphere of $290\pm55\mu$Jy, confirming at higher significance the excess seen at the star by other millimetre observations \citep{lestrade15,macgregor15,chavez16}. \epseri{} is known to be a magnetically active star with a bright chromosphere \citep[see][and references therein]{jeffers14}. \citet{gillett86} even tested whether emission from the chromosphere could explain the infrared excesses, but found the expected emission would be far too low at those wavelengths. At longer wavelengths, however, chromospheric emission can be bright enough to be detectable and this offers the best explanation for the detection of an excess above the photosphere at 7~mm with the Australia Telescope Compact Array \citep{macgregor15}. On the other hand, we know from shorter wavelength observations in the infrared that there is dust close to the star \citep{backman09,greaves14a}, that could be contributing to an unresolved excess. 

One way to determine the relative dominance of these is to compare to a star of a similar spectral type that is not known to have a significant debris disc. \citet{liseau15} have observed $\alpha$ Cen AB with ALMA, neither of which have shown any previous signs of a debris disc, and found a similar excess above the photosphere at long wavelengths, which they attribute to heated plasma from a chromosphere. $\alpha$ Cen B is a star of spectral type K1V \citep{torres06} and, therefore, we expect it to be similar enough to \epseri{} \citep[of spectral type K2V;][]{gray06} to estimate what the chromospheric contribution should be. \citet{liseau15} find that $\alpha$ Cen B has excesses above the photosphere of 40\% and 220\% at 870~$\mu$m and 3~mm respectively. Following \citet{chavez16}, we interpolate between these, assuming that the excess follows a power law, and find that the likely chromospheric contribution at 1.3~mm is $\sim$70\%. This would give a total stellar flux density of around 900~$\mu$Jy, which is slightly above our measured value suggesting that the entirety of excess we see is due to the chromosphere. This is different to the findings of \citet{chavez16} who find that, even including the chromosphere, the star makes up only about half of their measured central emission with $\sim$1.3~mJy being due to unresolved dust. Given the difference in beam sizes (11\arcsec{} with the LMT compared to our 1.4\arcsec) this gives extra weight to the hypothesis that there is dust in between the main ring and the star as any dust between about 2 and 20~AU would be unresolved by the LMT but resolved and, potentially, too spread out to be detected in these ALMA observations. Whilst \citet{backman09} do show evidence of dust at $\sim$3~AU, the quantity of dust is too low for this to contribute significantly at millimetre wavelengths.

\subsection{Azimuthal variations}
\label{sazi}

Previous observations in the sub-mm have noted a possible clumpy structure in the disc \citep{greaves98,greaves05,lestrade15,macgregor15}, although the Herschel \citep{greaves14a} and the LMT \citep{chavez16} observations find the disc to be smooth with a slight brightness asymmetry between the North and South ends of the disc (the South end being brighter). \citet{chavez16} show that there are a large number of galaxies to the East of the star's current location, which would have been behind the disc at the time of the previous observations and likely explains at least some of the clumps.

As noted in section \ref{smod}, our smooth models do leave some significant residuals. There are some residuals at the northern ansa (S1 and S2 in figure \ref{fmod}) and a point source to the northwest of the star (S3). Our further analysis of these shall consider the residuals from our best fit of run F (see figure \ref{ftworing}), although the residuals for the best fits of each run are roughly the same. The significant residual at the northern ansa peaks at 50$\pm$14~\ujypbm{} and is extended compared to the beam. A Gaussian fit gives an estimate of the total flux density within $2.76\arcsec\times2.40\arcsec$ to be 188$\pm$27~$\mu$Jy. The northwest point source is at a position angle of -63\degr{} and is located 16\arcsec{} from the star. If it is associated with the \epseri{} planetary system, its deprojected distance would therefore be 18\arcsec{} or 58~AU, just interior to the main ring. It has a flux density of 210$\pm$50~\ujypbm{} after correcting for the primary beam (this point is at the 28\% primary beam level, so we should be careful about over-interpreting this clump). 

\begin{figure}
	\centering
	\includegraphics[width=0.48\textwidth]{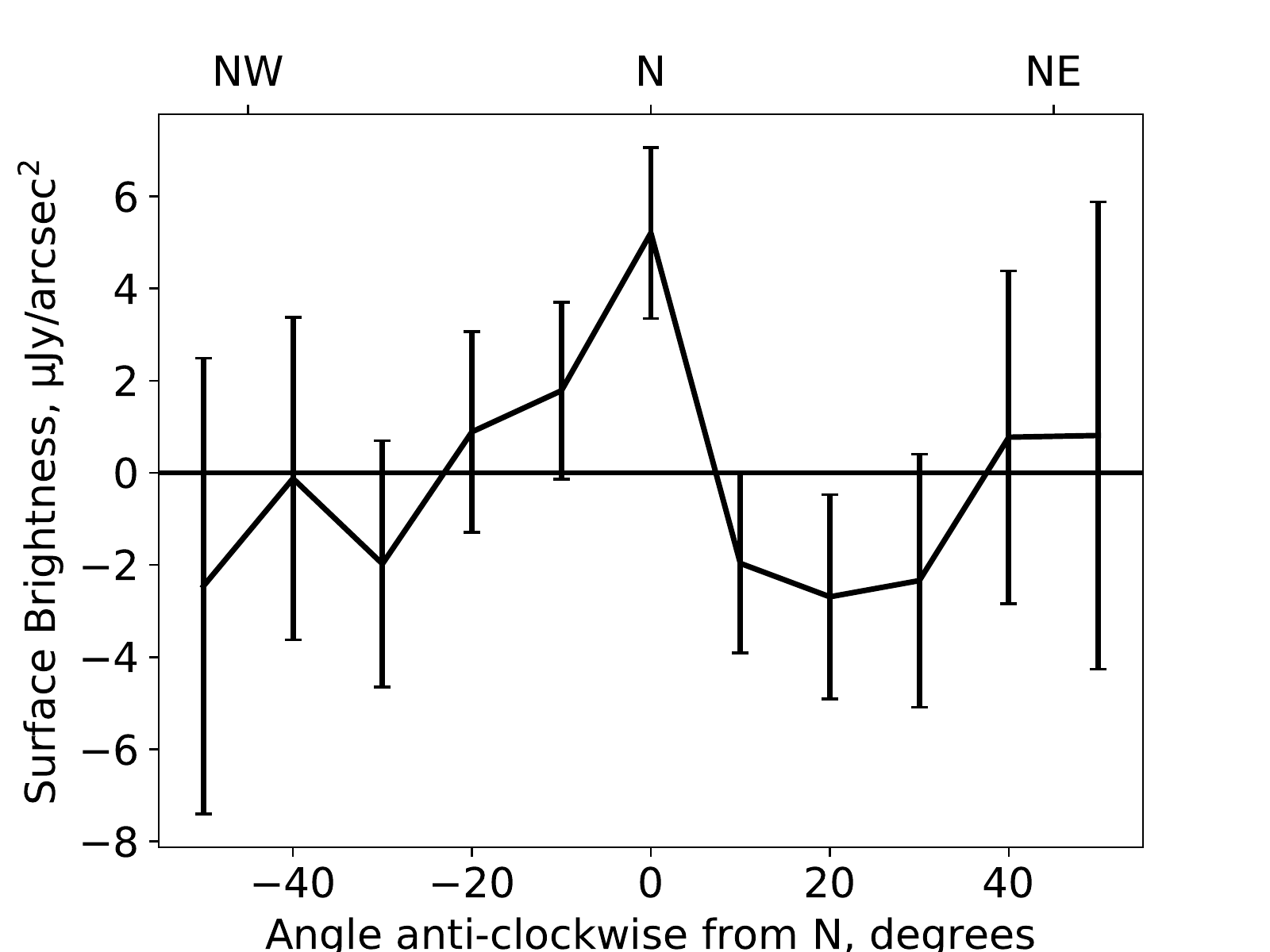}
	\caption{Azimuthal variation in the residuals after subtracting the best-fit model for run F. Each point is the mean of a 10\degr{} annular sector between deprojected radii of 15\arcsec{} and 24\arcsec{}.}
	\label{farc}
\end{figure}

Given the depth of this map, let us first consider whether any background galaxies are coincident with the ring. Galaxy number counts can be parametrised using a \citet{schechter76} function. The number of galaxies, $n$, per logarithmic bin can be found using
\begin{equation}
 n(F_\nu){\rm{d}}\log F_\nu = A\phi_\star\left(\frac{F_\nu}{S_\star}\right)^{\alpha +1}\exp\left(\frac{-F_\nu}{S_\star}\right)\ln10\,{\rm{d}}\log F_\nu
\end{equation}
where $A$ is the area of sky in deg$^2$ and $\phi_\star$, $S_\star$ and $\alpha$ are all constants of the equation. Many authors have investigated galaxy counts at sub-mm wavelengths. \citet{carniani15} have recently done this for ALMA observations at 1.3~mm and so their results are most useful to us. They find values of these constants of  $\phi_\star=1800\pm400$~deg$^{-2}$, $S_\star=1.7\pm0.2$~mJy and $\alpha=-2.08\pm0.11$. The significant residuals mentioned above are either within the main ring or just interior to it in the potential faint ring and have signal to noise ratios of at least 3.6$\sigma$. Considering this, we find the possibility of finding a galaxy between 15 and 24\arcsec{} (deprojected) from the star, $\pm$65\degr{} from North of at least 3.6$\sigma$ (here we are considering signal to noise ratio rather than sensitivity since the sensitivity varies across the image) to be 103\%. \citet{carniani15} show that physical models predict slightly lower galaxy counts than this Schechter function and, as they only go deep enough to detect galaxies brighter than 40~$\mu$Jy, they are not able to distinguish between these models, so the number we predict should probably be taken as an upper limit. Nonetheless, this does show that if any of the significant residuals in our image are real, it is likely that one is a galaxy.

The SCUBA \citep{greaves05}, IRAM \citep{lestrade15} and SCUBA-2 \citep{holland17} images all show tentative evidence for a `clump' to the northwest of the star that could be related to the point source that we detect, however, this does not appear in the deeper LMT image \citep{chavez16}. For comparison with the previous observations we also plot the azimuthal variation in the residuals for run F in figure \ref{farc}. Each point is the mean of a 10\degr{} annular sector between deprojected radii of 15\arcsec{} and 24\arcsec{} (note that this is slightly narrower than the analysis in previous papers since our higher resolution allows us to more accurately define the extent of the ring). This has been limited to sectors where all pixels are above the 20\% primary beam level as the noise increases drastically beyond this, which does mean that the northwest point source is not included. The uncertainties are calculated in the same way as for the radial profile (see section \ref{ssingle}). None of the azimuthal profiles of the previous observations show a rise to the north of the star, although comparing this azimuthal profile to that of \citet{macgregor15}, the rise we see directly north of the star does look similar to the rise they see $\sim$20\degr{} East of North, but the SMA observations were only taken roughly half a year prior to ours and so this would represent an infeasibly fast rotation rate (for comparison the Keplerian rate is 0.6\degr/yr at 70 AU). 

Follow-up observations will be necessary to determine whether any of the features seen here are real and associated with the planetary system.

\section{Conclusions}
In this paper we present the first observation of \epseri{} using ALMA. We focused our observation on the northern ansa of the main ring, 18\arcsec{} north of the star. Emission at the ansa of the ring is clearly detected at the 5$\sigma$ level, with a faint arc of emission from the ring stretching away from this. Whilst the signal-to-noise ratio along the arc is low, the area it covers compared to the beam size is large, enabling us to determine various parameters of the disc to a high degree of accuracy. We first consider three different single component models to fit the main belt. The best fits for all three have a FWHM between 11 and 13~AU, showing that the ring is narrow. For instance, assuming a sharp edged distribution, the observed disc emission is reproduced well with a model that extends from $63\pm1$~AU to $76\pm1$~AU. It is inclined at 34$\pm2$\degr{} from face on with a position angle of -4$\pm3$\degr{} measured east of north. The total flux density of the disc is found to be between 8 and 10~mJy, although caution should be taken here since this is based on a fit to only a small part of the ring and there is evidence for the southern side of the ring to be brighter, which is likely why this is discrepant with previous results.

The edges of the belt are found to be very steep. Considering both dynamical and observational constraints, if a planet is responsible for the location of the inner edge of the disc then it must have a $M_{pl}<1.3$~M$_J$ and $a>45$~AU. It is notable that the inner and outer edges of the Solar System's Kuiper belt line up with the 3:2 and 2:1 mean motion resonances of Neptune. If the same is true in the \epseri{} system, then we would expect a planet in the system to have a semi-major axis of 48~AU and a mass between 0.4 and 1.2~M$_J$.

We find tentative evidence for extended residual emission at the northern ansa and a point source to the North-West of the star, just interior to the disc. The latter of which could relate to clumps seen in previous observations, although there is a strong probability of it being due to a background galaxy. Further monitoring is necessary to determine whether these features are real and associated with the disc.

We find no strong evidence for emission around 20~AU in our data that has been previously reported in other observations. A narrow component at this distance would need to have $F_{1.34mm}<0.8$~mJy to not show up in our data whereas a flux density an order of magnitude higher would be possible if it were distributed more broadly. We do find tentative evidence for a faint, extra component just interior to the main disc, peaking at around 55~AU. Due to this observation being of such a small part of the ring, it is unclear whether this extra component is entirely due to the residual emission at the northern ansa or extends around the entire ring.

As with previous observations we find the stellar flux density to be higher than would be expected for a simple photosphere. This is consistent with a contribution from the chromosphere.

\section*{Acknowledgements}
The authors wish to thank Markus Janson for providing the observational limits from \citet{janson15}, Johan Olofsson and Claudio Caceres for help with fitting visibilities, and G. Fritz Benedict, Aaron Boley and Jacob White for useful discussions. The authors thank the referee for their helpful review. MB acknowledges support from a FONDECYT Postdoctoral Fellowship, project no. 3140479 and the Deutsche Forschungsgemeinschaft (DFG) through project Kr 2164/15-1. AJ and SC acknowledge financial support from the Millennium Nucleus RC130007 (Chilean Ministry of Economy). AJ acknowledges support from FONDECYT project
1130857, BASAL CATA PFB-06 and the Ministry for the Economy, Development, and Tourism's Programa Iniciativa Cient\'{i}fica Milenio through grant IC\,120009, awarded to the Millennium Institute of Astrophysics (MAS). MCW acknowledges the support of the European Union through ERC grant number 279973. GMK is supported by the Royal Society as a Royal Society University Research Fellow. JCA acknowledges support from PNP/CNES. This paper makes use of the following ALMA data: ADS/JAO.ALMA\#2013.1.00645.S. ALMA is a partnership of ESO (representing its member states), NSF (USA) and NINS (Japan), together with NRC (Canada), NSC and ASIAA (Taiwan), and KASI (Republic of Korea), in cooperation with the Republic of Chile. The Joint ALMA Observatory is operated by ESO, AUI/NRAO and NAOJ. The National Radio Astronomy Observatory is a facility of the National Science Foundation operated under cooperative agreement by Associated Universities, Inc. This research made use of Astropy, a community-developed core Python package for Astronomy \citep{astropy13}.

\bibliographystyle{mn2efix}
\bibliography{thesis}{}

\appendix
\section{Fitting to the visibilities}
\label{avfit}
The modelling in this paper is all done in the image plane. In theory, fitting a model to the dirty image is equivalent to fitting a model to gridded visibilities and should be a good approximation to fitting to the ungridded visibilities. In practice, each method will likely involve some averaging, which could plausibly bias the results. \citet{white16} model their ALMA data of Fomalhaut using both an image fitting method and visibility fitting method to compare the results. They find that in general the fitted parameters for the two different approaches are very similar accept for the total disc flux, which is 15\% lower in the image fitting approach. With this in mind, here we re-run run A but this time fitting to the visibilities.

To do this requires only some minor changes to the set-up. We average the visibilities to one channel for each of the four spectral windows. We create models at each of these four wavelengths, Fourier transform them and align them with the coordinates of the data. We then interpolate this image to provide the expected visibilities at the same $u$, $v$ coordinates as the data, calculating the $\chi^2$ values using the weights in the data. 

\begin{table}%
\begin{tabular}{cc}
Parameter & Fit \\ 
\hline
$R_{in}$ (AU) & 63.0$^{+0.6}_{-0.7}$ \\ 
$R_{out}$ (AU) &  76.3$^{+0.8}_{-0.7}$  \\ 
$\gamma$  & 1$^{+2}_{-2}$  \\ 
$F_{\nu}$ (mJy)  & 8.3$^{+0.6}_{-0.6}$ \\ 
$I$ (\degr) & 33$^{+2}_{-3}$  \\ 
$F_{cen}$ (mJy)  & 0.81$^{+0.04}_{-0.04}$ \\
$\Omega$ (\degr) & -4$^{+3}_{-3}$  \\
\end{tabular}
\caption{Fitting results for run A when fitting to the visibilities rather than the image.}
\label{tvis}
\end{table}

The results are shown in table \ref{tvis}. The differences between the results found when fitting to the visibilities and those found when fitting to the image (table \ref{tfit}) are small and all within the uncertainties. In particular, the total disc flux density is only 0.6\% higher. We, therefore, do not find the discrepancy seen by \citet{white16}. Considering the plots (their figures 3 and 5) of the residuals in \citet{white16}, it seems likely that the discrepancy they see is due to their image plane fit not successfully accounting for all of the flux in the ring rather than an inherent issue with fitting in the image plane.

\bsp
\end{document}